%% file: 0_main.tex
\documentclass[sigconf, screen=True, review = False]{acmart}
\usepackage[ruled,vlined,linesnumbered]{algorithm2e} % added for algorithms
\usepackage{enumitem} % added to remove space befroe itemize

\newcommand{\row}[3]{\multicolumn{1}{|p{2.6cm}|}{#1} & \multicolumn{1}{p{1.9cm}|}{#2} & \multicolumn{2}{p{3cm}|}{#3}}

\let\oldnl\nl% Store \nl in \oldnl
\newcommand{\nonl}{\renewcommand{\nl}{\let\nl\oldnl}}% Remove line number for one line

\AtBeginDocument{%
  \providecommand\BibTeX{{%
    \normalfont B\kern-0.5em{\scshape i\kern-0.25em b}\kern-0.8em\TeX}}}

\setcopyright{none}
\copyrightyear{}
\acmYear{}
\acmDOI{}

\acmConference[Preprint]{}{September}{2020}
\acmPrice{}
\acmISBN{}

\begin{document}

%%
%% The "title" command has an optional parameter,
%% allowing the author to define a "short title" to be used in page headers.
\title{A Framework for Checkpointing and Recovery of Hierarchical Cyber-Physical Systems}

%%
%% The "author" command and its associated commands are used to define
%% the authors and their affiliations.
%% Of note is the shared affiliation of the first two authors, and the
%% "authornote" and "authornotemark" commands
%% used to denote shared contribution to the research.
\author{Kaustubh Sridhar$\,^\Pi$, Radoslav Ivanov$\,^\Pi$, Vuk Lesi$\,^\lambda$, Marcio Juliato$\,^\lambda$, Manoj Sastry$\,^\lambda$, Lily Yang$\,^\lambda$, James Weimer$\,^\Pi$, Oleg Sokolsky$\,^\Pi$, Insup Lee$\,^\Pi$}
% \author{Kaustubh Sridhar}
% \email{ksridhar@seas.upenn.edu}
% \affiliation{PRECISE Center}
% \affiliation{University of Pennsylvania}

% \author{Radoslav Ivanov}
% % \email{rivanov@seas.upenn.edu}
\affiliation{$\;$\\$\,^\Pi\;$PRECISE Center, University of Pennsylvania $\;\;\;\;\;\;$ $\,^\lambda\;$Intel Labs}
% \affiliation{University of Pennsylvania}

% \author{Vuk Lesi, Marcio Juliato, Manoj Sastry, Lily Yang}
% \author{Vuk Lesi}
% \email{vuk.lesi@intel.com}

% \author{Marcio Juliato}
% \email{marcio.juliato@intel.com}

% \author{Manoj Sastry}
% \email{manoj.r.sastry@intel.com}

% \author{Lily Yang}
% \email{lily.l.yang@intel.com}
% \affiliation{Intel Labs}

% \author{James Weimer, Oleg Sokolsky, Insup Lee}
% \author{James Weimer}
% % \email{weimerj@seas.upenn.edu}

% \author{Oleg Sokolsky}
% % \email{sokolsky@seas.upenn.edu}

% \author{Insup Lee}
% \email{lee@seas.upenn.edu}
% \affiliation{PRECISE Center}
% \affiliation{University of Pennsylvania}

%%
%% By default, the full list of authors will be used in the page
%% headers. Often, this list is too long, and will overlap
%% other information printed in the page headers. This command allows
%% the author to define a more concise list
%% of authors' names for this purpose.
\renewcommand{\shortauthors}{Sridhar, K. et al.}

%%
%% The abstract is a short summary of the work to be presented in the
%% article.
\begin{abstract}
This paper tackles the problem of making complex resource-constrained cyber-physical systems (CPS) resilient to sensor anomalies. In particular, we present a framework for checkpointing and roll-forward recovery of state-estimates in nonlinear, hierarchical CPS with anomalous sensor data. We introduce three checkpointing paradigms for ensuring different levels of checkpointing consistency across the hierarchy. Our framework has algorithms implementing the consistent paradigm to perform accurate recovery in a time-efficient manner while managing the tradeoff with system resources and handling the interplay between diverse anomaly detection systems across the hierarchy. Further in this work, we detail bounds on the recovered state-estimate error, maximum tolerable anomaly duration and the accuracy-resource gap that results from the aforementioned tradeoff. We explore use-cases for our framework and evaluate it on a case study of a simulated ground robot to show that it scales to multiple hierarchies and performs better than an extended Kalman filter (EKF) that does not incorporate a checkpointing procedure during sensor anomalies. We conclude the work with a discussion on extending the proposed framework to distributed systems.
\end{abstract}

%%
%% The code below is generated by the tool at http://dl.acm.org/ccs.cfm.
%% Please copy and paste the code instead of the example below.
%%
\begin{CCSXML}
<ccs2012>
   <concept>
       <concept_id>10010520.10010553</concept_id>
       <concept_desc>Computer systems organization~Embedded and cyber-physical systems</concept_desc>
       <concept_significance>500</concept_significance>
       </concept>
   <concept>
       <concept_id>10010520.10010575</concept_id>
       <concept_desc>Computer systems organization~Dependable and fault-tolerant systems and networks</concept_desc>
       <concept_significance>500</concept_significance>
       </concept>
   <concept>
       <concept_id>10010520.10010575.10010577</concept_id>
       <concept_desc>Computer systems organization~Reliability</concept_desc>
       <concept_significance>300</concept_significance>
       </concept>
   <concept>
       <concept_id>10010520.10010553.10010554.10010556</concept_id>
       <concept_desc>Computer systems organization~Robotic control</concept_desc>
       <concept_significance>300</concept_significance>
       </concept>
 </ccs2012>
\end{CCSXML}

\ccsdesc[500]{Computer systems organization~Embedded and cyber-physical systems}
\ccsdesc[500]{Computer systems organization~Dependable and fault-tolerant systems and networks}
\ccsdesc[300]{Computer systems organization~Reliability}
\ccsdesc[300]{Computer systems organization~Robotic control}

%%
%% Keywords. The author(s) should pick words that accurately describe
%% the work being presented. Separate the keywords with commas.
\keywords{sensor anomaly resilience, state-estimate recovery, nonlinear and hierarchical cyber-physical systems, sensor fault tolerance, checkpointing}

%%
%% This command processes the author and affiliation and title
%% information and builds the first part of the formatted document.
\maketitle
\setlength{\textfloatsep}{10pt}
\input{1_Intro}

\input{2_background}
\input{3_problem_formulation}
\input{4_framework}

\input{5_errors}

\input{6_case_study}
\input{7_discussion}
\input{8_conclusion} %\KS{Change bibliographystyle from unsrt to ACM-Reference-Format before submission}

%%
%% The acknowledgments section is defined using the "acks" environment
%% (and NOT an unnumbered section). This ensures the proper
%% identification of the section in the article metadata, and the
%% consistent spelling of the heading.
%\begin{acks}

%\end{acks}

%%
%% The next two lines define the bibliography style to be used, and
%% the bibliography file.
\newpage
\bibliographystyle{ACM-Reference-Format}
\bibliography{sample-base}

%%
%% If your work has an appendix, this is the place to put it.
\newpage
$\;$
\newpage
\appendix
\input{9_appendix}

\end{document}

%% file: 1_Intro.tex
\section{Introduction and Related Work}
With the advent of cyber-physical systems (CPS) in critical day-to-day tasks from self-driving automobiles to medical devices, there is a heightened urgency to build systems resilient to sensor anomalies. 
%\RI{I think the idea was to call everything anomalies, not just attacks. This way, you wouldn't have to distinguish between faults and anomalies. But in the next paragraph, }
%\RI{\textbf{Done.} it is OK to say that anomalies may include faults or malicious attacks.}
%Sensors are a vital part of CPS that admit anomalies either directly \cite{camera_lidar_attks}, \cite{car_attks} or through non-invasive methods where the physical environment around a sensor is tampered with \cite{blackhat}, \cite{anti-lock}, \cite{GPS}. 
Sensors are a vital part of CPS that are easily susceptible to anomalies that can arise from hardware and software faults \cite{sensors}, \cite{sensors2}, changes in the physical environment around a sensor \cite{rado_ads_transient}, \cite{blackhat}, \cite{roboADS} or malicious attackers \cite{attack_paper}, \cite{GPS}. Some examples of faults include calibration issues \cite{sensors} and lost messages \cite{sensors2}. Examples of non-invasive anomalies from environment changes involve GPS obfuscation in tunnels \cite{rado_ads_transient}, obscured video cameras and lidars in adverse weather conditions \cite{blackhat}, interference between sensors such as infrared and lidar \cite{roboADS}, \textit{etc}. A couple of examples of malicious attacks are false-data injection attacks \cite{attack_paper} and GPS spoofing \cite{GPS}.
Ensuring the continued safe functioning of CPS under various kinds of sensor anomalies, however, is a challenging research problem.

Different solutions have been proposed for sensor anomaly resiliency of CPS \cite{roboADS}, \cite{anomaly1}, \cite{anomaly2}, \cite{attack_paper}, \cite{rado_ads_transient}, \cite{miroslav}
%,\cite{tabuada}
\cite{smt}, \cite{game}. Diverse anomaly detection systems (ADS) have been explored \cite{roboADS}, \cite{anomaly1}, \cite{anomaly2}, \cite{attack_paper}, \cite{rado_ads_transient}. Some ADS have been proposed for sensor anomaly detection \cite{anomaly1}, \cite{anomaly2}, \cite{attack_paper}, \cite{rado_ads_transient} whereas others are instrumental in finding actuator anomalies \cite{roboADS}. Building upon various ADS', resilient state estimators (RSE) \cite{miroslav} which solve a constrained minimization problem between sensor measurements and state evolution have been proposed. Other control methods resilient to anomalies based on SMT (Satisfiability Modulo Theory) \cite{smt} and game theory \cite{game} have also been delved into. 
%\KS{Should I replace SMT and game-theory methods as they are primarily for attack-resiliency and not anomalies?} \RI{I think it's OK to say that they are designed to detect attacks. We just use the term anomalies as an umbrella term.}

These methods are computationally intensive and don't account for hardware limitations. In practice, these algorithms will have to be implemented in platforms like embedded computers and micro-controllers with limited computational resources and system memory. On the other hand, vanilla state-estimation with Kalman Filters is not computationally taxing but also isn't resilient to sensor anomalies. To find a middle ground, 
%\RI{here maybe say ``cyber-physical checkpointing" to disambiguate from the original approach.}
cyber-physical checkpointing and roll-forward recovery was introduced in \cite{fanxin}.

Checkpointing and roll-forward recovery (RFR) \cite{fanxin} can be used in case of detected sensor anomalies (and hence incorrect state-estimates) in order to recover state-estimates from a periodically securely saved correct value (called a checkpoint). Between consecutive checkpoints, control inputs applied to actuators are also saved in secure memory. Recovery takes place by using these saved control inputs and "rolling" the system states forward from the checkpoint to the current time. Then, using the recovered state-estimate, we can generate a sensor anomaly resilient control command. 

It makes use of sensor ADS and can incorporate aspects of the RSE \cite{miroslav} and other sensor anomaly resilient control methods \cite{smt}, \cite{game} in the roll-forward procedure. This means that checkpointing and RFR can also be applied over existing resiliency methods as an extra layer of defense. 

In \cite{fanxin}, checkpointing and RFR is shown to be better than a Kalman Filter in prediction of state-estimate for linear time invariant (LTI) systems under sensor anomalies.
%\IL{\textbf{Done}. Is this last sentence true?  Need to elaborate.  What is attack-resilient control command?  Why is not used before the attack?}

Moreover, checkpointing and RFR can manage the tradeoff between system resources (memory and compute time) and accuracy of recovered state-estimate by not having to save every state-estimate, control input while also utilizing a shorter iterative "rolling" forward procedure. This in turn allows for its application to resource-constrained hardware.

Although a good first step, \cite{fanxin} does not address checkpointing and recovery for complex CPS with hierarchical distributed control architectures and nonlinear dynamics even though most CPS today, such as self-driving cars and diabetic implants, fall into that category \cite{vehicle}, \cite{ap}. 
%\IL{\textbf{Done}. Does the proposed work well for distributed systems?} 
There are two important challenges associated with checkpointing and RFR for nonlinear, hierarchical systems. 
%\IL{\textbf{Done}. State the next paragraph as an explicit list of challenges/issues.} 
%
%finding and enforcing consistency is (1); 1 sentence for system complexity introduces challenges: interplay bw ADS and preserve accuracy of state estimation while preserving tradeoff. More intro into tradeoff (Maybe added no-linearity) ; put numbers (1) and more concise
(1) Hierarchical systems necessitate algorithms to snapshot and locate consistent checkpoints with coordination between sub-systems. (2) With increased complexity from non-linearities and the hierarchy, we need to direct attention to preserving the accuracy of recovered state-estimates while managing the tradeoff with system resources, coordinating amongst sub-systems with diverse ADS outputs and ensuring timely-communication between sub-systems with a time-efficient roll-forward strategy.

We tackle the first challenge by introducing three checkpointing paradigms for checkpointing with different levels of consistency. We detail a coordinating algorithm scalable to multiple hierarchies, each operating at different frequencies.

We address
%\RI{maybe say ``address", to avoid repetition}
the second challenge with an analysis of the tradeoff between accuracy and system resources and demonstrate the ability to manage this tradeoff by tuning checkpointing frequency. We propose a RFR algorithm that performs time-efficient recovery with minimal checkpoint usage while allowing for the interplay between sub-systems with different ADS (specific or generic). 
%This paper describes our framework that consists of 3 checkpointing paradigms and algorithms to solve \RI{\textbf{Done.} Which are? If you make the previous paragraph more structured and clear, then here you can just say which specific parts you are addressing. Saying that you are addressing all issues is the best way to get rejected. :)}
%the issues of consistent checkpointing, increased accuracy \textit{vs} resource utilization, time-efficient roll-forward recovery and adaptation to diverse ADS outputs in nonlinear and hierarchical systems. 
%\RI{This can be part of the previous paragraph. Looks a bit weird here}
We also discuss resolutions for the issues of communication security and minimal lag in extending our framework to distributed systems. 

%Our framework consists of 3 checkpointing paradigms and algorithms to both, perform checkpointing with variable frequency employing the consistent paradigm and, perform roll-forward recovery in all attacked sub-systems, regardless of the type of ADS, in a time-efficient manner.
%At the end of every time-step, our proposed framework uses the recovered state-estimates to calculate sensor-attack resilient control inputs. \IL{\textbf{Done}. What is the framework? what are its various methods?}

%This paper describes our framework that addresses all these issues and present a spectrum of methods for checkpointing and roll-forward recovery of hierarchical systems \IL{Due to what?}\IL{what are relations between different methods and "to...frequencies, ...outputs, etc.} corresponding to differing checkpointing frequencies, diverse ADS outputs and varied checkpoint usage.

%\IL{\textbf{Done}. First, state what are good properties to guarantee and why. And then what our framework guarantees.} 
Further, there is a need to guarantee bounds on the difference between recovered state-estimate and ground truth (\textit{a.k.a.} recovered state-estimate error) and quantify the maximum permissible duration of application of the above algorithms. Additionally, investigating the tradeoff between system resources and accuracy leads to questions about the accuracy-resource gap (which is the difference between an accuracy-optimal state-estimate recovered with most-frequent checkpointing and one from some arbitrary checkpointing frequency).

In this paper, under assumptions on bounded noise, jacobian and certain non-linearity, we guarantee bounds on recovered state-estimate error regardless of linear or nonlinear dynamics. We analyze its variation under different checkpointing frequencies which in turn leads to an analysis in tradeoffs between accuracy and system resources and bounds on the accuracy-resource gap. Also, given maximum admissible error, we present a bound on the maximum tolerable duration of anomalies in sensor data.

We evaluate the proposed framework on a case-study of a simulated ground robot with one outer loop and two inner loops. The case study demonstrates scalability of the framework to multiple hierarchies and that the proposed framework lowers estimation error (when given anomalous sensor data) significantly when compared to an extended Kalman filter estimator in every loop.

In summary, this paper makes the following contributions.
\begin{itemize}[noitemsep,topsep=0pt]
    \item A framework for checkpointing and roll-forward recovery (RFR) of nonlinear, hierarchical systems (Section \ref{framework_sec})
    \item A theoretical analysis of bound on recovered state-estimate error, maximum tolerable anomaly duration and tradeoff between checkpointing frequency, accuracy and system resources. (Section \ref{err_sec})
    \item A numerical validation on a simulated ground robot that demonstrates scalability and efficacy. (Section \ref{case_study_sec})
\end{itemize}
%\item A discussion on use-cases (Section \ref{case_study_sec}) and further extension of the framework to distributed systems (Section \ref{discussion_sec})

%% file: 2_background.tex
\section{Background and Preliminaries} \label{rel_sec}
In this section, we mention the background information on cyber and physical checkpointing and recovery. 
%\RI{\textbf{Done below}. Maybe add 1-2 sentences describing the original checkpointing idea at a high level, just for context.}
\subsection{Cyber state checkpointing and rollback recovery}
Cyber checkpointing and rollback recovery \cite{cyber} is an important fault-tolerance technique for long-running applications on distributed computing systems (\textit{e.g.} distributed optimization). It lets the application restart from the last saved snapshot (or checkpoint) of its state rather than restarting from the beginning.

\begin{figure}[t]
    \centering
    \includegraphics[width=0.8\linewidth]{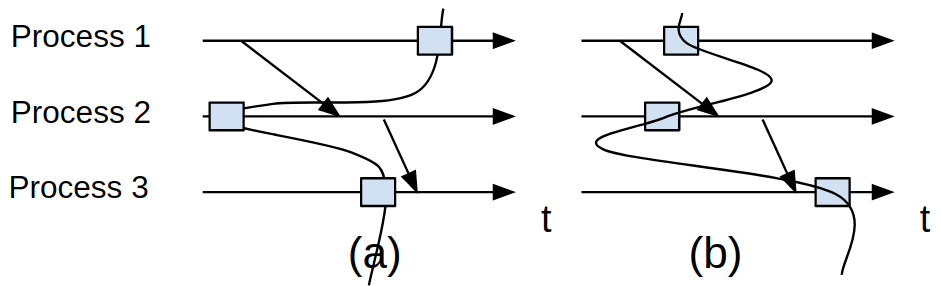}
    \setlength{\abovecaptionskip}{5pt}
    \setlength{\belowcaptionskip}{-5pt}
    \caption{(a) Consistent \& (b) inconsistent cyber checkpoint}
    \label{cyber-ckpt}
\end{figure}

Cyber state recovery takes place by rolling-back to a checkpoint and restarting a task. Consistent checkpoints are preferred as they prevent the domino effect (\textit{i.e.} multiple rollbacks to find a state with sufficient information to restart the task)
%\RI{This is lacking context. It's not clear what the domino effect is. If you explain this at the top, maybe it will be fine here.} 
\cite{cyber}. A consistent cyber state checkpoint is a periodically saved consistent cyber state. 
\begin{definition}
A consistent cyber state is a cyber state in which no message is recorded as received in one process and as not yet sent in another process \cite{cyber}. \label{cyber-def}
\end{definition}
Figure \ref{cyber-ckpt} shows examples of consistent and inconsistent cyber checkpoints. Note that a cyber state with all elements saved at the same time is consistent since it satisfies Definition \ref{cyber-def}. %Thus Equation \eqref{cyber_state} also represents a consistent cyber state.
\subsection{Physical state-estimate checkpointing and roll-forward recovery (RFR)}
In \cite{fanxin}, the following simple linear time invariant (LTI) system is considered
\begin{align}
    x_{k+1} &= A x_k + B u_k + w_k \label{lti1}\\
    y_k &= C x_k + v_k \label{lti2}
\end{align}
where 
$x_k \in \mathbb{R}^n$, $u_k \in \mathbb{R}^m$, $y_k \in \mathbb{R}^p$, $w_k \sim \mathcal{N}(0, Q)$ and $v_k \sim \mathcal{N}(0,R)$. State-estimates (denoted $\hat{x}_k$) are produced by a Kalman Filter. 
%Physical state estimate checkpointing and roll-forward recovery \cite{fanxin}, with inspiration drawn from cyber checkpointing, was introduced to recover state-estimates of linear time invariant (LTI) systems with one feedback control loop \RI{\textbf{Done}. What is ``single-loop"?} by repeated prediction (\textit{a.k.a.} "rolling" forward) of state-estimate from a saved consistent checkpoint of the state-estimate.

The concept of checkpointing and physical state recovery for the simple LTI system \eqref{lti1}, \eqref{lti2} was introduced in \cite{fanxin}. Physical state-estimate recovery occurs by rolling state-estimate elements forward to current time from a consistent checkpoint. Checkpoints are values of state-estimate saved periodically in stable secure memory.

\begin{figure}[t]
    \centering
    \includegraphics[width=0.75\linewidth]{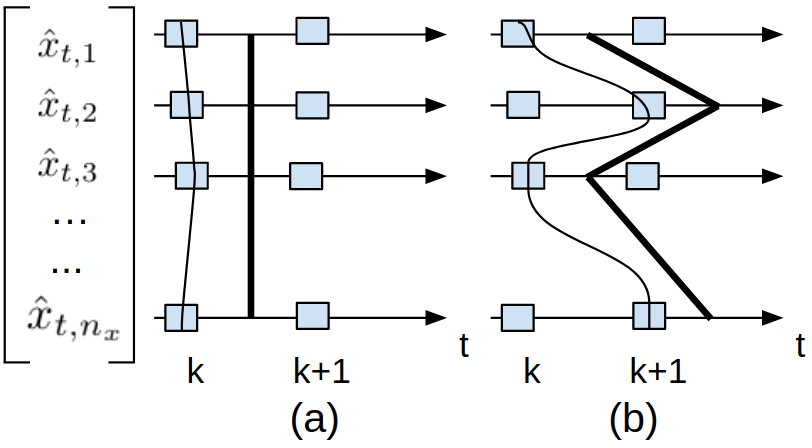}
    \setlength{\abovecaptionskip}{5pt}
    \setlength{\belowcaptionskip}{-5pt}
    \caption{(a) Consistent and (b) inconsistent physical state-estimate checkpoints for a LTI system.}
    \label{cps-ckpt-fanxin}
\end{figure}

\begin{definition} \label{fanxin-def}
Consistent physical state-estimate of a LTI system such as the one in \eqref{lti1}, \eqref{lti2} is a physical state-estimate which has all of its elements' values from the same time-instant \cite{fanxin}.
\end{definition} %\KS{I've been forced to use simple system instead of sub-system so that I can use extend this Defn to Defn 4.1. \textbf{Check}}
Figure \ref{cps-ckpt-fanxin} shows examples of consistent and inconsistent physical state-estimate checkpoints for LTI system via boxes with curved line passing through. These boxes represent values at or just before checkpointing time that are saved as the checkpoint. Note that the notion of "the same time" can be modified to allow a bounded precision between clocks used to timestamp local states. For simplicity, this paper assumes the existence of globally consistent time \cite{Kopetz2011}. 
%\KS{simple system here too and in Fig 2}
%\IL{\textbf{Done}. Added the last note.  reference Kopetz's book chapter 3 on global time - need to make global time granularity is less than half of smallest period under synchronous execution}

%As a background to our proposed framework, we summarize roll-forward recovery for single-loop LTI systems from \cite{fanxin} here. The authors of \cite{fanxin}
%\RI{\textbf{Check}. Probably move the system definition to the beginning of the section for more context.} \KS{move to before and don't talk about single-loop/subsytems until the end of this subsection.}
%consider the following LTI system, 
%\begin{align}
%    x_{k+1} = A x_k + B u_k + w_k\\
%    y_k = C x_k + v_k
%\end{align}
%where $x_k \in \mathbb{R}^n, u_k \in \mathbb{R}^m, y_k \in \mathbb{R}^p, w_k \sim \matcal{N}(0, Q)$ and $v_k \sim \mathcal{N}(0,R)$. State-estimates are produced by a Kalman Filter (denoted $\hat{x}_k$). 

Next, the algorithm in \cite{fanxin} assumes a specific ADS that tells the controller (at time $k$) that the first $q$ elements of the state-estimate have been affected. These affected elements are denoted $\hat{x}_{k,(1,q)}$ and healthy elements are denoted $\hat{x}_{k,(q+1,n)}$. 

When $q$ is at least one, RFR takes place. Physical state-estimate (at time $k$) is recovered from a checkpoint of state-estimate that was created $N$ time steps behind current time (denoted $\Bar{x}_{k-N}$), using control inputs saved between the checkpoint time and current time, with repeated use of the Kalman Filter's predict step ($x_{k+1} = A x_k + B u_k$) as follows,
\begin{align}
    \hat{x}^R_k &= A^N \Bar{x}_{k-N} + \sum_{i=1}^{N} A^{i-1}Bu_{k-i}.
\end{align}

%\RI{Using ``they" too much is informal. It is important what the algorithm is, not what the authors did per se.}
Then, the $q$ affected elements of state-estimate are updated with roll-forward recovered value as follows,
\begin{align}
    \hat{x}_{k,(1,q)} &= \hat{x}^R_{k,(1,q)}.
\end{align}
Lastly, this updated state estimate is used to generate control inputs that are sent to an actuator. We reiterate that this work on checkpointing and recovery in \cite{fanxin} is limited in application to a simple LTI system with an assumed specific ADS. 

We present a more general problem with a nonlinear hierarchical system that consist of 2 feedback control loops (\textit{a.k.a.} sub-systems) in the following section. Each of the sub-systems may have a specific or generic ADS (Section \ref{ADS_sec}). Our framework, introduced in Section \ref{framework_sec}, is applicable to this generalized problem and can be extended to distributed systems under conditions discussed in Section \ref{discussion_sec}.

%\RI{\textbf{Removed and check}. This paragraph is out of place here.}

%Our framework saves a snapshot of a consistent state-estimate across the hierarchy during checkpointing with each sub-system having a different loop rate. It also searches for and finds the most recent consistent checkpoint (across sub-systems) before the start of the anomalous data for roll-forward recovery. It handles increased roll-forward time due to the nonlinear system dynamics with a time-efficient roll-forward procedure. It works alongside both specific and generic ADS in every sub-system. It can manage the trade-off between higher accuracy of recovered state-estimate and available system resources by tuning checkpointing frequency. 

%% file: 3_problem_formulation.tex
\section{Problem Formulation} \label{pf_sec}
In this section we present the hierarchical system model and measurement model, anomaly model, estimator, controllers, anomaly detection systems, cyber-physical system model and objective.
\begin{figure}[t]
\centering
\includegraphics[width=\linewidth]{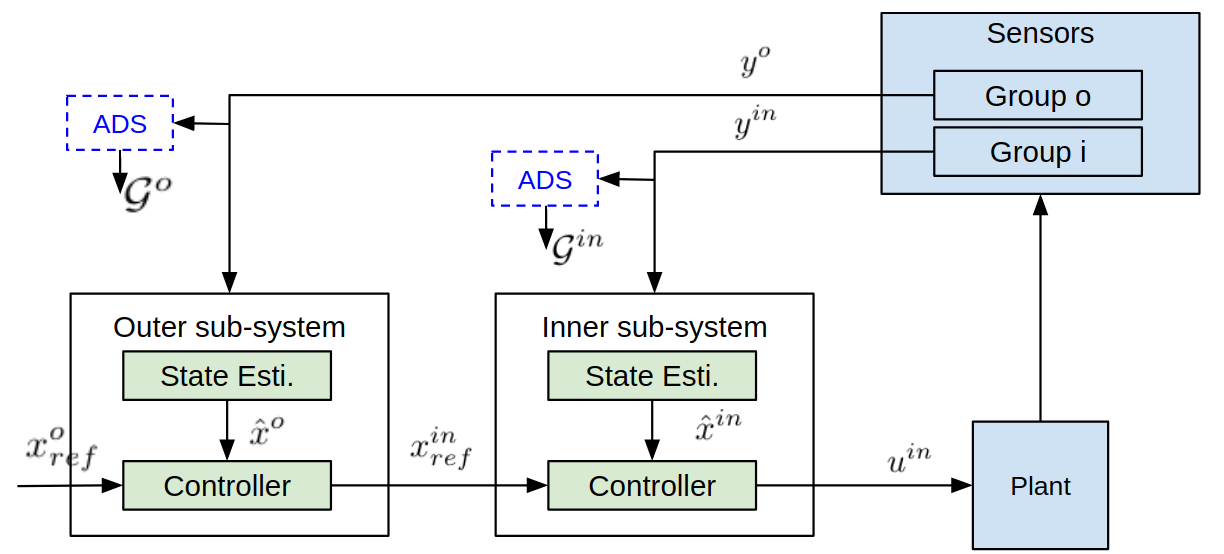}
\setlength{\abovecaptionskip}{5pt}
\setlength{\belowcaptionskip}{-5pt}
\caption{Block diagram of the hierarchical system}
\label{block_diag_sys}
\end{figure}
\subsection{Physical System and Measurement Model} \label{physical_sec}
We assume that the physical system is a collection of the inner and outer sub-systems (or loops) as shown in Figure \ref{block_diag_sys} and represented with superscript $j \in \{in,o\} $. We have discrete-time nonlinear sub-system dynamics and measurement models as follows
\begin{align}
    x^j_{k+1} &= f^j(x^j_k,u^j_k) + \omega^j_k, \;\;\;\; x^j_{0} \sim \mathcal{N}(\mu^j_{0}, \Sigma^j), \;\;\; \omega^j_k \sim \mathcal{N}(0, Q^j)
    \label{sys} \\
    y^j_{k} &= g^j(x^j_k,u^j_k) + \gamma^j_k, \;\;\;\;\; \gamma^j_k \sim \mathcal{N}(0, R^j)
    \label{meas}
\end{align}
where $x^j_k \in \mathbb{R}^{n_x^j}$, $y^j_k \in \mathbb{R}^{n_y^j}$ and $u^j_k \in \mathbb{R}^{n_u^oj}$ are the physical state vector, measurement vector and control input (at time $k$) of the $j$th sub-system. We have process and measurement covariance matrices $Q^j, R^j$ for the $j$th subsystem.

\subsection{State Estimation} \label{EKF_sec}
In each of the $(in)$ and $(o)$ sub-systems, we assume an estimator that calculates state-estimate with a predict step and improves it with an update step. Some examples are the RSE \cite{miroslav} \& the extended Kalman filter (EKF) which can be represented, in the $j$th sub-system, with a predict step \eqref{pred_gain}, gain calculation \eqref{pred_gain} \& update step \eqref{upd}, %\JW{I am concerned that you're setting the wrong tone of the paper by introducing all this irrelevant background math that has NOTHING to do with our contribution.  The problem is that reviewers are going to see KF/EKF and think this is trivial.  why can't we just introduce a function that provides a state estimate, rather than a state estimator?}
\begin{align}
    \hat{x}^j_{k+1|k} &= \text{predict}^j(\hat{x}_k^j, u_k^j)  \;\;\;\;\; K^j_k = \text{gain}^j(\hat{x}_k^j, Q^j, R^j) \label{pred_gain}\\
    \hat{x}^j_{k+1} &= \text{update}^j(\hat{x}^j_{k+1|k}, K^j_k, {y^j_k}') \label{upd}
\end{align}

\subsection{Controllers} \label{control_sec}
\subsubsection{Reference Trajectory}
We assume that reference trajectory's state (at time $k$) denoted $x^o_{k, ref}$ is provided to the outer loop. The outer loop generates control $u^o_{k}$ which is transformed by map $T: \mathbb{R}^{n^o_u} \to  \mathbb{R}^{n^{in}_x}$ to obtain inner loop reference value $x^{in}_{k, ref} = T(u^o_{k})$.
\subsubsection{Control}
We assume that each loop uses a controller abstracted as $u^j_{k} = h^j(\hat{x}^j_{k}, x^j_{k, ref}), \;\;\; j \in \{in,o\}$.
%\begin{align}
%    u^j_{k} &= h^j(\hat{x}^j_{k}, x^j_{k, ref}), \;\;\; j \in \{in,o\}. \label{u^{in}}
%\end{align}
We also assume that each loop $j$ runs at a frequency $\mu^j$. %\OS{\textbf{Done}. I think we need to say something about rates of the two subsystems here.  We know that each loop has its own rate, but I do not see it introduced until Section 4.  Especially because Figures 5 and 6 give an impression that the rates are the same.}
Some examples of controllers that can be used are PID \cite{fanxin}, neural network \cite{NN_CPS}, dynamic inversion \cite{clever}, model predictive  \cite{mpc_car} controllers, \textit{etc}.
%We also assume bounds on aforementioned control inputs as given above. If computed value of control input exceeds bounds, the control inputs takes its value equal to the violated bound.

\subsection{Anomaly Model} \label{attk_sec}

We assume that sensor anomalies lead to change in the readings of a subset of sensors (represented as $\mathcal{S}^{a}$), which results in the following generalized model:
\begin{eqnarray}
    &{y^j_{k}}' = g^j({x^j_k}',u^j_k) + \gamma^j_k + \Gamma^j y^{j,a}_{k}, \;\;\; j = in,o
    \label{attk_meas}
\end{eqnarray}
where $y^{j,a}_k \in \mathbb{R}^{n_y^j}$ is the anomalous addition (at time $k$) to the sensor readings of the $j$th sub-system. We also have sensor selection matrices $\Gamma^{j} = diag(\eta^{j}_1, ..., \eta^{j}_{n_y^{j}})$ where $\eta^j_i = 1$ if $i$th sensor of sub-system $j$ is in $\mathcal{S}^{a}$. This results in anomalous measurement vectors ${y^j_{k}}' \in \mathbb{R}^{n_y^j}$ and affected states ${x^j_{k}}' \in \mathbb{R}^{n_x^j}$ (differs by $'$ from healthy values). As difficulty in recovering accurate state-estimates increases with anomaly duration, we primarily concern ourselves with transient/intermittent anomalies such as GPS signal loss in tunnels \cite{rado_ads_transient}, lidar affected by reflective surfaces, fog or infrared light \cite{blackhat}, cameras impaired by harsh lights or rain \cite{roboADS}, \textit{etc.}. 
%\KS{recheck this}
%\KS{Should I bring section 6.1 here?}
%\RI{\textbf{Done but check}. talk about transient anomalies, which are the obvious class of anomalies handled by our approach} \KS{As difficulty in recovering ; make forward reference}

\subsection{Anomaly Detection System (ADS)} \label{ADS_sec}
% Marcio Juliato Comment: I would recommend using Intrusion Detection System (IDS) throughout the document instead of Attack Detection System (ADS)
%Different ADS like the  $\chi^2$ detector \cite{attack_paper}, \cite{attack_paper_2}, the sequential probability ratio test (SPRT) detector \cite{attack_paper_drones}, sensor-fusion based detector \cite{sfd}, \cite{fuseADS1} and others \cite{surveyofsurveys} have been proposed. 
Most ADS fall into two broad categories - a specific
%\RI{Are these names common in the literature? The ``generic" detector is probably better known as a change detector. I am not sure if there is a word for the specific one -- maybe Jim will know.} 
detector and a generic detector. The specific detector (\textit{e.g.} deep neural network detector \cite{anomaly1}, $\chi^2$ detector \cite{attack_paper}, sensor-fusion detector \cite{rado_ads_transient}) provides a list of sensors with anomalies whereas the generic detector (\textit{e.g.} fuzzy logic detector \cite{anomaly2}) can only inform the system about the presence of some anomalous data without specifying which sensors have anomalies. We assume an ADS in each sub-system of the hierarchy and abstract an ADS of sub-system $j$ as,
\begin{align}
    \mathcal{G}^j_k &= f_{ADS}({y^j_k}', ..., {y^j_{k-\mathcal{T}^j+1}}',k) \label{bool_vec}
\end{align}
where the detection function $f_{ADS}:\mathbb{R}^{\mathcal{T}^j (n_y^j)} \to \{0,1\}^{m}$ takes as input a window of measurements of size $\mathcal{T}^j$ (detection time for sub-system $j$) and current time-instant $k$. It outputs either a Boolean vector $\mathcal{G}^j_k \in \{0,1\}^{n_y^j}$ whose $i$th element takes a value of 1 when the $i$th sensor has anomalous data (for a specific detector) or a simple Boolean $\mathcal{G}^j_k \in \{0,1\}$ representing the presence of some sensors with anomalous data (for a generic detector). Thus the same abstraction can be used for both specific and generic detectors and each sub-system may have either of the two.

Further, some ADS such as the sensor-fusion based detector \cite{rado_ads_transient} require that at least some $\zeta>0$ sensors are healthy at time $k$ for detection. Others like the $\chi^2$ detector \cite{attack_paper} assume prior knowledge on sensor performance for detection. Thus, the extent of protection delivered by our framework and its performance in recovering state-estimates depends on the detection capability of the ADS.
%\KS{ADS is in each sub-system}

%\OS{\textbf{Done}. This sentence is vague.  Our framework is better than any IDS because it does more.  Maybe: "Thus, protection (efficiency, performance?) delivered by our framework depends on the detection capability of ADS."} 
%\MJ{Should we address detection latency in this subsection?}
% Marcio Juliato Comment: Should we address detection latency in this subsection?

\subsection{Cyber-Physical System} \label{CPS_sec}
The CPS is a combination of the physical elements (physical system, sensors, actuators) and software elements (state estimators, ADS, controllers) described in the previous section. The system's physical state (denoted $x_{k, p} \in \mathbb{R}^{n_x^{in}+n_x^o}$) contains the physical states of each sub-system. The system's cyber state (denoted $x_{k, c} \in \{0,1\}^{m^{in}+m^o}$) contains the Boolean output sent by the ADS to the sub-systems. The CPS state (denoted $x_{k, CPS} \in \mathbb{R}^{n_x^{in}+n_x^o} \times \{0,1\}^{m^{in}+m^o}$) is a combination of the system's physical and cyber states. We have physical, cyber and CPS state as follows,
\begin{align}
    x_{k, p} &= \begin{bmatrix} x^{in}_k \\ x^o_k \end{bmatrix} \;\;\;
    x_{k, c} = \begin{bmatrix} \mathcal{G}^{in}_k \\ \mathcal{G}^o_k \end{bmatrix} \;\;\; 
    x_{k, CPS} = \begin{bmatrix} x_{k, p} \\ x_{k, c} \end{bmatrix} \label{all_states}.
\end{align}
%The CPS state-estimate is obtained from a combination of the extended Kalman filters and the ADS as given in subsections \ref{EKF_sec}, \ref{ADS_sec}.\RI{Wait, the ADS is not used in the estimation yet. This is part of your solution, not part of the system in Fig. 1. I'd remove this sentence.} %The physical state-estimate is represented as $\hat{x}_{k, p} &= \begin{bmatrix} \hat{x}^{in}_k & \hat{x}^o_k \end{bmatrix}^T$.

\subsection{Objective}
%Our objective is to generate sensor attack-resilient control commands $u^{in}_k, u^o_{k}$ at time $k$ using historical data of state-estimates and control inputs by performing cyber-physical checkpointing and roll-forward recovery. Our additional objective is to ensure that \textbf{recovered state-estimate error assuming attack duration and communication delay is bounded.} \IL{What do you mean by "communication remain bounded?"} 
Our primary objective is to recover CPS state-estimates at time $k$ when facing sensor anomalies using the checkpointing and RFR framework while working within the system's hardware constraints. This includes striving for a balance between accuracy and resource utilization. Then, we use recovered state-estimates to generate sensor anomaly resilient control commands $u^{in}_k, u^o_{k}$. Our secondary objective is to ensure that the recovered state-estimate error remains within acceptable bounds by halting application of proposed framework after maximum tolerable anomaly duration.

%% file: 4_framework.tex
\section{Framework for Checkpointing and Recovery} \label{framework_sec}
\begin{figure}[t]
\centering
\includegraphics[width=\linewidth]{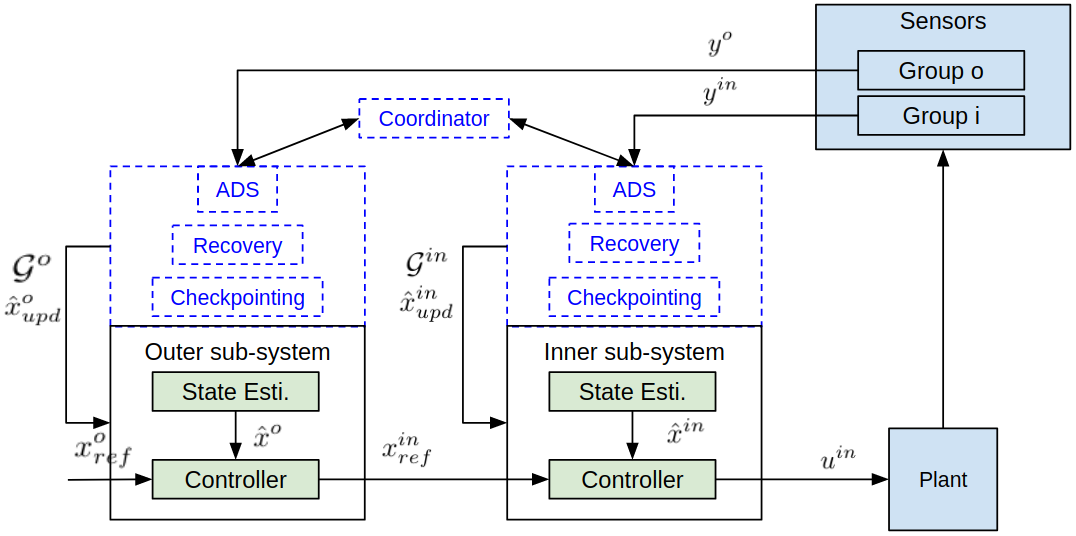}
\setlength{\abovecaptionskip}{3pt}
\setlength{\belowcaptionskip}{-5pt}
\caption{Block diagram of the framework and ADS (in blue) integrated into the hierarchical system}
\label{block_diag_sys_frame}
\end{figure}

In this section, we propose our consistent checkpointing and RFR framework for hierarchical systems, comprised of Algorithms \ref{fmwk_algo}, \ref{rf_algo}, \ref{cood_algo}, that integrates into the hierarchical system as shown via components called ADS, "recovery", "checkpointing" and "coordinator" in Figure \ref{block_diag_sys_frame}. 
We begin this section by comparing three checkpointing paradigms and their analogous frameworks in Section \ref{prelims}. We propose using the consistent paradigm and associated consistent framework. Then, we explain the checkpointing and RFR protocols of Algorithms \ref{fmwk_algo}, \ref{rf_algo} in Sections \ref{fmwk_ckpt_sec}, \ref{rf_sec} and the role of the coordinator (Algorithm \ref{cood_algo}) in Section \ref{cood_sec}. %We also introduce tradeoffs in checkpointing frequency, system memory and accuracy in Section \ref{tradeoff_sec}.

\subsection{Checkpointing Paradigms} \label{prelims}
\begin{figure}[t]
    \centering
    \includegraphics[width=\linewidth]{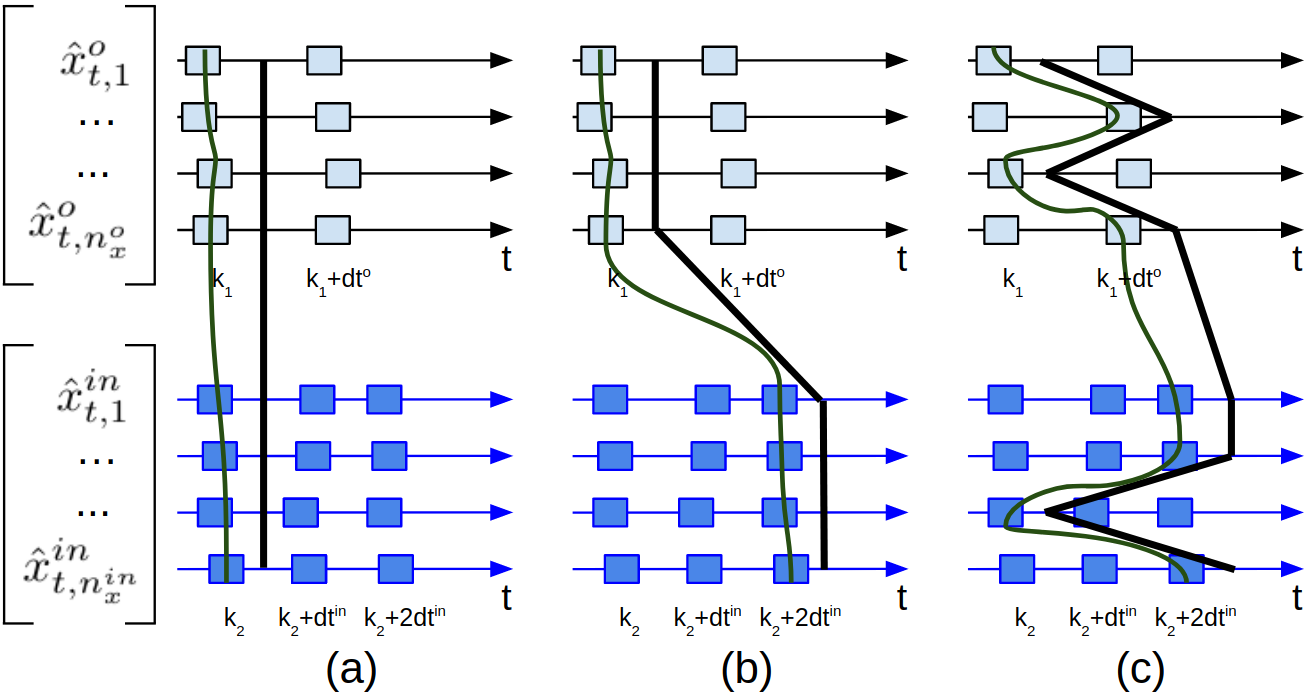}
    \setlength{\abovecaptionskip}{5pt}
    \setlength{\belowcaptionskip}{-5pt}
    \caption{(a) Consistent,(b) partly inconsistent \& (c) fully inconsistent physical state-estimate checkpoints for hierarchical CPS. Values saved at/ just before checkpointing time.}
    \label{cps-ckpt}
\end{figure}

For dealing with hierarchical systems, we extend Definition \ref{fanxin-def} to,
\begin{definition}
Consistent physical state-estimate for the hierarchical CPS is a same time-instant combination of consistent physical state-estimates of every sub-system (which has all of it's elements from the same time-instant).
\end{definition}
An inconsistent physical state-estimate of the hierarchical CPS can be either partly or fully inconsistent (see Figure \ref{cps-ckpt}(b) and (c)). Partly inconsistent physical state estimates in the CPS are combinations of consistent physical state-estimates of inner and outer subsystems at different time-instants (see Figure \ref{cps-ckpt}(b)). Fully inconsistent physical state-estimates are combinations of inconsistent physical state-estimates of inner and outer subsystems (see Figure \ref{cps-ckpt}(c)). Checkpoints are physical state-estimates of the CPS saved every user-specified time-interval in stable secure memory. 

%Figure \ref{cps-ckpt} shows examples of consistent and inconsistent physical state-estimate checkpoints for the CPS.

\begin{figure}[t]
    \centering
    \includegraphics[width=\linewidth]{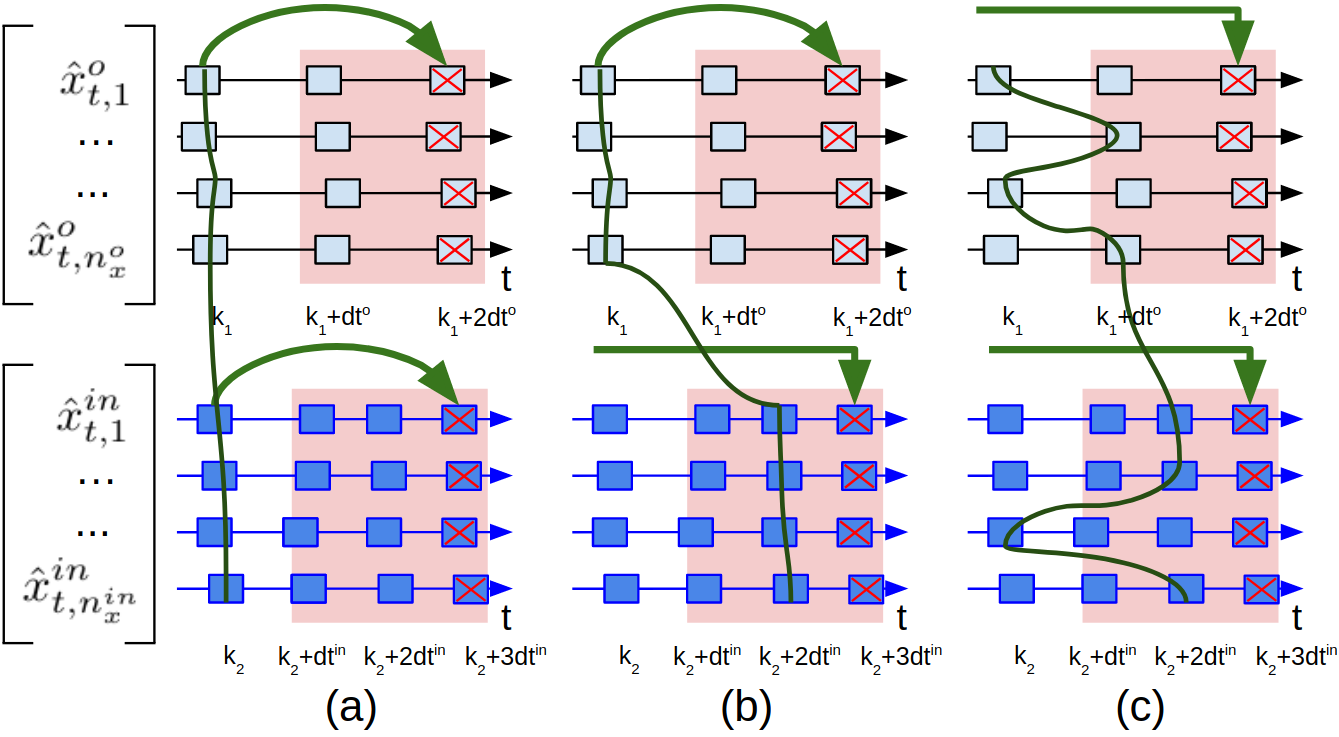}
    \setlength{\abovecaptionskip}{5pt}
    \setlength{\belowcaptionskip}{-5pt}
    \caption{(a) Proposed consistent framework and alternates- (b) partially inconsistent, (c) fully inconsistent frameworks when anomalous sensor data starts with red shaded region and is detected at values with red cross marks. RFR (green arrows) uses latest checkpoint before start of the anomaly.}
    \label{issue}
\end{figure}

The distinctions between the three types of checkpoints is important when considering possible alternates to the proposed framework. We define them below. 
\begin{definition}
Consistent framework: proposed framework with recovery from consistent physical state-estimate checkpoints of CPS.
\end{definition} 
\begin{definition}
Partially inconsistent framework: alternate framework where physical state-estimate checkpoints consistent in each sub-system but not in the hierarchy are used for recovery.
\end{definition} 
\begin{definition}
Inconsistent framework: alternate framework where fully inconsistent physical state-estimate checkpoints of CPS are used for recovery.
\end{definition} 
These three frameworks are shown in Figure \ref{issue} (a), (b) \& (c), where all elements of outer and inner state estimates have been detected as anomalous at times $k_1+2\text{dt}^{o}$ and $k_2+3\text{dt}^{in}$ because of sensor anomalies that started at $k_1+\text{dt}^{o}$ and $k_2+\text{dt}^{in}$ respectively.

In framework (a), we have RFR from consistent checkpoints at $k_1$ (for the outer loop) and $k_2$ (for the inner loop) depicted by curved green arrows. 

In framework (b), sub-system $(o)$ with checkpoint at $k_1$ can recover from said checkpoint. On the other hand, sub-system $(in)$ cannot recover from its failed checkpoint at $k_2+2\text{dt}^{in}$. Instead, the recovery (depicted by right-angled green arrow) has to begin from the checkpoint behind the most-recent checkpoint which is not shown in the figure. If this checkpoint is distant to the current time, we will have different roll-forward times for each sub-system. In practice, this could lead to delays in computing control inputs and undesirable (or even unstable) behaviour.

In framework (c), some elements of both subsystems $(in), (o)$ have checkpoints in the anomalous period and so we cannot have recovery from said checkpoints for those elements. Instead, for those elements, the recovery (depicted by right-angled green arrows) has to begin from the checkpoint behind the most-recent checkpoint which is again not shown in the figure. Thus, similar to (b), this could result in undesirable behaviour of the CPS. 

Although all the above three frameworks can find a conceivable use in hierarchical CPS, we employ only the first consistent framework throughout this work. %\IL{\textbf{Done}. Should state three frameworks are possible, but we pick the first one.}
This is because, we see that, in the situation depicted in Figure \ref{issue}, the proposed consistent framework fares better than alternate inconsistent frameworks (b) and (c). Further, identifying most-recent checkpoint outside anomaly duration for each and every sub-system in (b), and for each and every element in (c), requires increased computational resources and time as compared to the proposed framework (a).

Thus, to avoid both undesirable behaviour and additional overhead in resource utilization, we proceed with the consistent framework rather than the alternate inconsistent frameworks.

\subsection{Consistent framework for the hierarchical system and its checkpointing protocol} \label{fmwk_ckpt_sec}
Algorithm \ref{fmwk_algo} presents the consistent checkpointing procedure and initiates the RFR when anomalous data is received in every sub-system $j$.

In Line 1, the algorithm sets the loop rate $\mu^j$ and initializes state-estimate $\hat{x}_0^j$. In lines 2-18, we have a loop running at the specified loop rate which terminates after a preset time-instant T or when anomaly exceeds maximum tolerable duration (Line 17). At time-instant $k$ in Line 3, Algorithm 1 begins by receiving a coordinator Boolean $c_k^j$ from the \textbf{consistent checkpointing coordination protocol} of Algorithm \ref{cood_algo}. If this coordinator Boolean evaluates to 1, a checkpointing procedure (Lines 13-17) may be initiated downstream. Before that, the sub-system estimates state $\hat{x}^j_k$ and receives estimator gain $K_k^j$ in Line 4 via predict, gain calculation and update steps. In Line 5, the ADS takes in a window of measurements (of length detection time $\mathcal{T}^j_k$ when $k \geq \mathcal{T}^j_k$ or else of length $k$) and gives out either a Boolean vector (for a specific detector) or a simple Boolean (for a generic detector) as previously discussed in Section \ref{ADS_sec}. The algorithm also has access to value of detection time $\mathcal{T}^j_k$. 

In Line 6, it calculates the 'anomaly detected' Boolean as either a union of the Boolean elements of $\mathcal{G}^j_{k}$ for a specific detector or the simple Boolean $\mathcal{G}^j_{k}$ itself for a generic detector. If an anomaly was detected (\textit{i.e.}, the 'anomaly detected' Boolean is 'True'), then it calls the RFR algorithm given by Algorithm \ref{rf_algo} (Lines 7-9). Suppose an anomaly was detected, then after obtaining the recovered state-estimate, it calculates an anomaly-resilient control value in Line 10. If no anomaly was detected, a standard control value is calculated in Line 10. 

Lastly, control inputs are securely saved in an array (Line 11). Further, if no anomaly was detected and the coordinator Boolean received at the start evaluates to 1, a checkpoint of cyber-physical state-estimate (\textit{i.e.}, physical state-estimate and ADS outputs) is appended to an array of checkpoints (Lines 12,13). The time of checkpointing is also appended to an array (Line 14). Both these arrays are securely saved in Line 15. If the current anomaly's duration exceeds the maximum tolerable anomaly duration, then it exits to a safe stop (Line 17) or else it repeats this process.
%\OS{\textbf{I think Done}. My concern with algorithm 1 is that it, conceptually, runs for every $k$.  However, the sub-systems do not necessarily run at every k}
\input{algos/algo1}
\subsection{Roll-forward recovery (RFR) protocol for the hierarchical system} \label{rf_sec}
We explain Algorithm \ref{rf_algo} which presents the RFR protocol in this subsection. Once a sub-system receives anomalous sensor data, Algorithm \ref{rf_algo} is triggered and it proceeds to find the latest consistent checkpoint across all sub-systems (in Line 2) via the \textbf{most-recent consistent checkpoint time protocol} of Algorithm \ref{cood_algo}.
\input{algos/algo2}

Then, when the anomaly is first detected at time-instant $k$, \textit{i.e.} when no anomaly was detected at previous time-instant $k-1$, it calculates roll-forward estimate $\hat{x}^{j, R}_k$ from the most-recent physical state-estimate checkpoint at time-instant $k_1$ (given by $\Bar{x}^j_{k_1}$) via repeated prediction (Line 4) using securely retrieved control inputs (Line 1). If an extended Kalman Filter (Appendix \ref{EKF_sec_app}) is used for estimation, then predict$^j(x,u)=f^j(x,u)$. For every time-instant after first-detection, it calculates roll-forward estimate from the previous value (Line 6). This strategy is time-efficient and precludes delays from repeated prediction with non-linear dynamics.

Second, provided a specific detector is used (Line 8), it identifies all the state-estimate elements, at time-instant $k$, which depend on anomalous sensor measurements. This takes places by dimension change of Boolean vector of anomalous sensors $\mathcal{G}^{j}_{k}$ to Boolean vector of state-estimates' elements that depend on anomalous sensors $\mathrm{G}^{j}_{k}$ as shown in Line 9. Then, it updates elements of state-estimate that depend on anomalous measurements with corresponding element of roll-forward estimate in Lines 10-12. It does not change healthy elements.
If suppose a generic detector was used instead (Line 13), it simply updates the complete state-estimate with roll-forward estimate (Line 14).
\input{algos/algo3}

\subsection{Coordinator} \label{cood_sec} 
Algorithm \ref{cood_algo}, which details the functioning of the "coordinator" depicted in Figure \ref{block_diag_sys_frame}, is described in this subsection. It comprises two protocols. The first one, \textbf{consistent checkpointing coordination protocol} (Lines 1-9), consists of a loop operating at frequency $\max\{\mu^o, \mu^{in}\} = 1 / \min\{dt^o, dt^{in}\}$ until preset time-instant T.
%\OS{\textbf{I think Done}. I am not convinced that this is enough, given that rates of the two sub-systems may be different.  Do we assume harmonic rates?} 
The coordinator Booleans of $(in)$ and $(o)$ sub-systems ($c_{k}^{in}, c_{k}^{o}$) take values $1,1$ at a frequency of $\mu = 1/dt^c$ (Lines 3,4). In between consecutive values of 1, the coordinator Booleans take values $0,0$ (Lines 5-7). The protocol sends these coordinator Booleans to each sub-system at every time-instant (Line 8). This checkpointing frequency is chosen by the user in Line 2 by managing the trade-off between accuracy and available system resources. This is elaborated on in Section \ref{tradeoff_sec}.

The second protocol in Algorithm \ref{cood_algo} is the \textbf{most-recent consistent checkpoint time protocol} on Lines 10,11. It finds the most-recent consistent checkpoint time across sub-systems that is not within any detection period through Procedure \textbf{last common element outside of ADS detection times}(save\_times$^{in}$, save\_times$^{o}$, $\mathcal{T}_k^{in}$, $\mathcal{T}_k^{o}$) $\to$ $k_1$ by finding $k_1$ such that $k - k_1 > \max(\mathcal{T}^{in}_k, \mathcal{T}^o_k)$.
In the next section, we analyze the sources of error and the trade-off between accuracy of recovered state-estimate and systems resources (Subsection \ref{tradeoff_sec}). The trade-off analysis leads us to a meaningful choice of checkpointing frequency $\mu$ to achieve high accuracy in resource-constrained systems.

%% file: algos/algo1.tex
\begin{algorithm}[t]
\DontPrintSemicolon
\SetAlgoLined
\SetKwInOut{Input}{input}\SetKwInOut{Output}{output}
\Input{sensor measurements: ${y_k^{j}}'$, reference trajectory: $x^j_{k, ref}$, maximum tolerable anomaly duration: $\mathcal{T}^{j}_{\text{max}}$}
\Output{anomaly-resilient control commands: $u_k^{j}$}
\textbf{initialize:} loop rate = $\mu^j = \frac{1}{dt^j}$, $\hat{x}_{0}^j$\;
\For{k = 1 to T in time steps of $dt^j$}{
    $c_k^j$ $\gets$ short wait for Boolean from \textbf{consistent checkpointing coordination}() in coordinator\;
    $\hat{x}^{j}_{k}, K^j_k \gets \text{update}^j( \text{predict}^j (\hat{x}^{j}_{k-1}, u_{k-1}^j), \text{gain}^j(\hat{x}_{k-1}^j, Q^j, R^j), {y_{k}^{j}}')$\;
    $\mathcal{G}^{j}_k$, $\mathcal{T}^{j}_k$ $\gets$ $f_{ADS}({y^j_k}', ..., {y^j_{k-\mathcal{T}^j_k+1}}',k)$\;
    anomaly\_detected$_k^j$ = $\begin{cases} \bigcup\limits_{i=1}^{n_y^j} \mathcal{G}^{j}_{k,i}, \text{ for specific detector} \\ \mathcal{G}^{j}_{k}, \text{ for generic detector} \end{cases}$\;
    \If{anomaly\_detected$_k^j$ = True}{
        $\hat{x}^{j}_{k}$ $\gets$ \textbf{roll-forward recovery $\left( \hat{x}^{j}_{k}, K^j_k, \mathcal{G}^{j}_k, \mathcal{T}_{k}^{j} \right)$}
    }
    $u_{k}^j$ $\gets$  $h^j \left( \hat{x}_{k}^j, x^j_{k, ref} \right)$\;
    controls.append $\left( u_{k}^j \right)$; \textbf{secure\_save} controls\;
    \If{anomaly\_detected$_{k}^j$ = False \textbf{and} $c_k^j$ = True}
    { %\tcc*{Checkpoint Creation in Sub-system $j$}
        checkpoints.append $\left( \hat{x}^{j}_{k}, \mathcal{G}^{j}_{k} \right)$\;
        save\_times.append($k$)\;
        \textbf{secure\_save} checkpoints, save\_times
    }
    \lIf{anomaly duration $ > \mathcal{T}^{j}_{\text{max}}$}{\textbf{exit to safe stop}}
}
\caption{Consistent Checkpointing, Roll-forward Recovery and Sensor-anomaly Resilient Control Framework on Sub-system $j \in \{in, o\}$}
\label{fmwk_algo}
\end{algorithm}

%% file: algos/algo2.tex
\begin{algorithm}[t]
\DontPrintSemicolon
\SetAlgoLined
\SetKwInOut{Input}{input}\SetKwInOut{Output}{output}
\Input{state-estimate $\hat{x}_k^j$, estimator gain matrix $K^j_k$, ADS Boolean(s) $\mathcal{G}^{j}_k$ and detection time $\mathcal{T}_{k}^{j}$}
\Output{recovered / updated state-estimate $\hat{x}_{k}^j$}

\textbf{secure\_retrieve:} checkpoints, save\_times, controls\;
$k_1$ $\gets$ \textbf{most-recent consistent checkpoint time from coordinator}($\mathcal{T}_k^{in}$, $\mathcal{T}_k^{o}$)\;
\uIf{anomaly\_detected$_{k-1}^j$ = False}{
    $\hat{x}^{j, R}_{k} \gets \text{predict}^j(...\text{predict}^j ( \Bar{x}^{j}_{\text{checkpoint at }k_1}, u_{k_1}^j).., u_{k-1}^j)$\;
}
\Else{
    $\hat{x}^{j, R}_{k}$ $\gets$ $\text{predict}^j ( \hat{x}^{R}_{k-1}, u_{k-1}^j)$
}
\uIf{specific detector}{
    elements of $\hat{x}_k^j$ that depend on anomalous sensors= $\mathrm{G}^j_k := K^j_k \mathcal{G}^j_k$ with non-zero elements replaced by $\mathbf{1}$\;
    \For{index $i$ \textbf{where} ${G}^j_{k, i}==1$}
    {
        $\hat{x}^{j}_{k, i} = \hat{x}^{j, R}_{k, i}$
    }
}
\ElseIf{generic detector}{
     $\hat{x}^{j}_{k} = \hat{x}^{R}_{k}$
}
\caption{Roll-forward Recovery Function on Sub-system $j \in \{in, o\}$}
\label{rf_algo}
\end{algorithm}

%% file: algos/algo3.tex
\begin{algorithm}[t]
\DontPrintSemicolon
\SetAlgoLined
\SetKwInOut{Input}{input}\SetKwInOut{Output}{output}
\nonl\hspace{-0.32 cm} \textbf{def} consistent checkpointing coordination\textbf{:}\;
\Input{none}
\Output{coordinator Booleans $c_k^{in}, c_k^o$}
\textbf{initialize:} $\mu=\frac{1}{dt^c}$ $\gets$ \textbf{choose optimal checkpointing frequency}(system resources, desired accuracy)\;
\For{k = 0 to T in time steps of $\min\{dt^o, dt^{in}\}$}{
    \uIf{k \% ($dt^c$) = 0}{
    $c_k^{in}, c_k^o$ $\gets$ 1, 1
    }
    \Else{
    $c_k^{in}, c_k^o$ $\gets$ 0, 0
    }
    send $c_k^{in}, c_k^o$ to sub-systems $(in), (o)$
}
\nonl\;
\nonl\hspace{-0.32 cm} \textbf{def} most-recent consistent checkpoint time\textbf{:}\;
\Input{ADS detection times $\mathcal{T}_k^{in}$, $\mathcal{T}_k^{o}$}
\Output{consistent checkpoint time $k_1$}
\textbf{secure\_retrieve:} save\_times$^{in}$, save\_times$^{o}$\;
$k_1$ $\gets$ \textbf{last common element outside of ADS detection times}(save\_times$^{in}$, save\_times$^{o}$, $\mathcal{T}_k^{in}$, $\mathcal{T}_k^{o}$)

\caption{Coordinator between sub-systems $(in)$ \& $(o)$}
\label{cood_algo}
\end{algorithm}

%% file: 5_errors.tex
\section{Error and Tradeoff analysis} \label{err_sec} 
When a sub-system receives anomalous sensor data, 3 types of errors are observed: detection error, recovered state-estimate error (RSEE), standard estimation error. In this section we analyze the cause of the 3 errors and propose bounds on the RSEE and corresponding maximum tolerable anomaly duration. We scrutinize the trade-off between checkpointing frequency, recovered state estimate accuracy \& system resources and propose bounds on the accuracy-resource gap that this trade-off begets.

\subsection{Causes of errors}
During detection time in each sub-system, sensors may have anomalies and yet no anomalies would be detected. The error induced in state-estimates' elements that depend on anomalous sensors (when compared to ground truth) during this time-period is called detection error. In this work, we assume a small detection time ensuring bounded detection error. Further, as previously mentioned in Section \ref{ADS_sec}, the protection delivered by our framework is dependent on the strength of the ADS. That is, we cannot account for errors induced by anomalies that are never detected by the ADS. We leave an exploration of the solution for the same to future work.
%If a sneaky attack has a large detection-time, then, the corresponding detection error has to be small for the sneaky attack to have had avoided detection for so long. That is, there is an inherent trade-off between detection time and detection error.
$\;\;\;\;\;$ 

After the anomaly has been detected, RFR is triggered. During recovery, we update the state estimate at particular elements (for specific ADS) or completely (for generic ADS). The difference in updated state-estimate and ground truth gives rise to the RSEE. The difference is engendered when repeated predict steps cannot take process noise into account leading to increasing accumulation of drift of recovered value from ground truth. We present bounds on this error in the next section.

%\KS{I believe the table might have a lot of redundant info. \textbf{Ask Profs Oleg/Insup to check!} Maybe move to appendix?}

Further, for a specific ADS, some state-estimate elements are obtained directly from the estimator. Since our sensor measurements are noisy, we have the standard estimation error for these elements. We will also have the standard estimation error when there are no anomalies in a sub-system and all state-estimate elements are obtained directly from the estimator.
We characterize solutions/ assumptions for safe functioning of CPS in a case-by-case analysis of different anomaly situations in each subsystem in Appendix \ref{case-by-case-analysis}.

\subsection{Bound on recovered state-estimate error} \label{bound_subsec}
Let us suppose that some $q$ elements of consistent physical state-estimate of sub-system $j$ at time $k$ (denoted $\hat{x}^{j}_{k, (1,q)} \in \mathbb{R}^q$) are updated by the RFR process. We can have $q \in {1, ..., n_x^j}$. Then, we have recovered error for subsystem $j$ as 
\begin{align}
    e^{j}_{k, \ Rec} &= x^{j}_{k,(1,q)} - \hat{x}^{j,R}_{k,(1,q)} \label{pred_error_eq}.
\end{align}
where $x^{j}_k$ is the ground truth. We can also represent the $n_x^j-q$ healthy elements (at time $k$, of sub-system $j$) as $\hat{x}^{j}_{k,(q+1,n_x^j)} \in \mathbb{R}^{n_x^j-q}$ with associated estimation error as $e^{j}_{k, \ Est} = x^{j}_{k,(q+1,n_x^j)} - \hat{x}^{j}_{k,(q+1,n_x^j)}$.
%\begin{align}
%    e^{j}_{k, \ Est} = x^{j}_{k,(q+1,n_x^j)} - \hat{x}^{j}_{k,(q+1,n_x^j)} %= : \delta^j_{k,(q+1,n_x^j)}.
%\end{align}
Let $e^{j}_{k, \ Est} = \delta^j_{k,(q+1,n_x^j)}$ where $\delta^j_{k}=x_k^j - \hat{x}_k^j$ can be obtained via a method similar to that employed in \cite{fanxin} by solving the following,
\begin{align}
    y_k^j &= g(x_k^j, u_k^j) + \gamma_k^j \;\;\;\;\; \text{ and } \;\;\;\;\; y_k^j = g(\hat{x}_k^j, u_k^j) \label{process_delta}
\end{align}
If the underlying estimator (EKF, KF, RSE \textit{etc.}), when using healthy sensor measurements, has bounded estimation error at the start ($k=0$) and so at time $k$, \textit{i.e.},
\begin{align}
        &\delta^j_k \preceq \epsilon^j_{\delta_k} \;\; \forall k\geq0 \label{err_ass_0}
    \end{align}
then, we have bounded estimation error (in healthy elements),
\begin{align}
        e^{j}_{k, \ Est} &\preceq \epsilon^j_{\delta_k, (q+1, n_x^j)} \label{ee_b}
    \end{align}
where matrices $A \preceq B \implies A$ is element-wise less than/equal to $B$.

\begin{proposition} \label{prop_2}
    The RSEE is bounded as follows,
    \begin{align}
        e^{j}_{k+1, \ Rec} &\preceq \left[  |(\mathbf{A}^j)^{k-k_1+1}| \epsilon^j_{\delta_{k_1}} + \sum_{l=k_1}^{k} |(\mathbf{A}^j)^{k-l+1}| \epsilon^j_{\omega_k} + \Bar{\Phi}^{j,R}  \right]_{(1,q)}  \label{pe_b} \\
        &=: \mathcal{B}(k, k_1) \label{Bound}
    \end{align}
    where we assume Inequality \eqref{err_ass_0} at checkpoint time $k_1$ and also assume (for $j=in,o$),
    \begin{align}
        &A^j_k \preceq \mathbf{A}^j, \omega^j_k \preceq \epsilon^j_{\omega_k} \label{err_ass_1}\\
        &\Phi(x^j_k,\hat{x}^{j,R}_k, u^j_k, ..., x^j_{k_1},\Bar{x}^{j}_{k_1}, u^j_{k_1}) \preceq \Bar{\Phi}^{j,R} \;\;\; \forall k \ \geq0 \label{err_ass_2}
    \end{align}
    where $\Phi(...)$ represents the nonlinear terms from each successive Taylor series expansion of $\hat{x}^{j,R}_{k+1} = f^j (...f^j (\Bar{x}^{j}_{k_1}, u^j_{k_1})..., u^j_{k})$ about $\hat{x}^{j,R}_k$
\end{proposition}
Proposition \ref{prop_2} states that RSEE is bounded by the RHS of Inequality \eqref{pe_b} if we assume bounds on Jacobian $A^j_k$, process noise $\omega^j_k$, underlying estimation error at checkpoint time $\delta_{k_1}$ and nonlinear term from successive Taylor expansions of $\Phi(x^j_k,\hat{x}^{j,R}_k, u^j_k, ...)$. Proof of Proposition \ref{prop_2} can be found in Appendix \ref{Proof_1}.

\begin{corollary} \label{lin_corollary}
If we have LTI system dynamics, \textit{i.e.}, $f^j(x^j_k,u^j_k) = Ax^j_k + Bu^j_k$, nonlinear term is removed to obtain the following bound on RSEE
\begin{align}
    e^{j}_{k+1, \ Rec} &\preceq \left[  |(\mathbf{A}^j)^{k-k_1+1}| \epsilon^j_{\delta_k} + \sum_{l=k_1}^{k} |(\mathbf{A}^j)^{k-l+1}| \epsilon^j_{\omega_k}\right]_{(1,q)} \label{lin_bound_pred}
\end{align}
Equations \eqref{lin_bound_pred}, \eqref{ee_b} (RSSE and estimation error bounds for the particular case of LTI dynamics) were verified previously in \cite{fanxin}.
\end{corollary}
The Corollary states that the RSEE is bounded by the RHS of Inequality \eqref{lin_bound_pred} when LTI system is considered and bounds on Jacobian $A^j_k$, process noise $\omega^j_k$ and underlying estimation error at checkpoint time $\delta_{k_1}$ are assumed as before.

Further, from Equations \eqref{pe_b} and \eqref{lin_bound_pred}, we see that RSEE bound may increase with increase in $k-k_1$ (current time \textit{minus} most-recent checkpoint time), \textit{i.e.} as time for which anomaly exists increases. This results from the dead-reckoning based RFR that accumulates drift with passage of time.

\subsection{Maximum Tolerable Anomaly Duration} \label{max_attack_duration_sec}
\begin{proposition} \label{prop_3}
    The maximum tolerable anomaly duration $T_{\max}^j$ for an anomaly that started at time $s$ in sub-system $j$ can obtained as
    \begin{align}
        \text{arg}\min\limits_{T^j_{\max}} \ \left\lvert \mathcal{B}(T_{\max}^j + s, \mathcal{F}^j(s, \Delta, \mu)) - \mathbf{E}^j \right\rvert \label{T_min_prob}
    \end{align}
    where we assume maximum permissible error for sub-system $j$ as $\mathbf{E}^j$. We also assume a relationship between latest consistent checkpoint time before anomaly $(k_1)$, start time of anomaly $(s)$, minimum time between anomalies $(\Delta)$ and checkpointing frequency $(\mu)$ as $k_1 = \mathcal{F}^j(s, \Delta, \mu)$. Note that $\mathcal{B}(.,.)$ is given by \eqref{pe_b}, \eqref{Bound}. 
\end{proposition}
Proposition \ref{prop_3} states that we can find the maximum anomaly duration by solving for the time-instant when the bound on RSEE is equal to the maximum permissible error. The proof of this proposition is given in Appendix \ref{Proof_3}.

\subsection{Choosing checkpointing frequency: trade-off between accuracy and available system resources} \label{tradeoff_sec} 
%\KS{Should 4.5 be combined with 5.4?}\RI{I think so. You can end this section with a small segueway saying that the next section will analyze all the sources of error and the various trade-offs.}
There is an inherent trade-off between recovered state-estimate accuracy and available system memory that can be manipulated by changing the checkpointing frequency. We will more often have a checkpoint close to the start of an anomaly if checkpointing frequency is high. This results in fewer predict steps which in turn reduces the accumulated error (or increases accuracy) for a recovered state-estimate. Also, a high checkpointing frequency involves rapid usage of system memory. This can be an impediment for micro-controllers and embedded computers with memory limitations. Besides, more predict steps increases compute time which can place resource-constrained hardware in a predicament wherein they are unable to complete RFR within one time-step and so engender some undesirable (or even unstable) plant behaviour. 

With respect to accuracy, the optimal checkpointing frequency is given by $\mu^* \geq \max(\mu^{in}, \mu^{o})$. Such a value of $\mu^*$ ensures checkpoint creation at every time-step across the hierarchy which always guarantees a consistent checkpoint just before the start of any anomaly. This further leads to the least number of predict steps in the recovery process and the highest possible accuracy. We prove bounds for the difference between an accuracy-optimal recovered state-estimate and one that is obtained with some arbitrary checkpointing frequency $\mu$ below. Please note that we call this difference the accuracy-resource gap since it is a byproduct of the trade-off in accuracy and computational resources.

\subsubsection{\textbf{Bound on Accuracy-Resource Gap}} \label{optimal_bound}
The accuracy-resource gap (as discussed above) is the difference between a state-estimate recovered when checkpointing frequency is $\mu^* \geq \max(\mu^{in}, \mu^{o})$ (\textit{i.e.} a checkpoint is created every time-instant in both $(in)$ \& $(o)$) and one that is obtained with some arbitrary checkpointing frequency $\mu$. Now, we propose a bound on the accuracy-resource gap as follows.
\begin{proposition} \label{prop_4}
    The bound on accuracy-resource gap at time-instant $k$, given by $\mathcal{P}^j_k = \hat{x}_{k,(1,q)}^{j,R} \Big\rvert_{\mu^*} - \hat{x}_{k,(1,q)}^{j,R} \Big\rvert_{\mu}$, in sub-system $j$ with an anomaly that started at time-instant $s$ is as follows
    \begin{align}
        \mathcal{P}^j \leq \mathcal{B}(k, \mathcal{F}^j(s, \Delta, \mu)) - \mathcal{B}(k, s - 1)
    \end{align}
    where $\mathcal{F}^j(s, \Delta, \mu^*) = s-1$, $\mathcal{F}^j(.,.,.)$ is defined in Proposition \ref{prop_3} and $\mathcal{B}(.,.)$ is given by \eqref{pe_b}, \eqref{Bound}.
\end{proposition}
Proposition \ref{prop_4} states that the accuracy-resource gap is bounded by the difference in RSEE bounds for a framework with some arbitrary checkpointing frequency and one with optimal checkpointing frequency. Proof of Proposition \ref{prop_4} can be found in Apppendix \ref{Proof_4}.

%% file: 6_case_study.tex
\section{Use Cases and Evaluation} \label{case_study_sec} 
In this section we describe the types of anomalies that the framework can handle and evaluate its working on a case-study of a numerically simulated differential drive ground robot.
\subsection{When can the framework be used?} 
%\KS{Should this really be moved up? I believe this location is better as it allows me to talk about the framework stopping after max anomaly duration}
We intend for our framework to be used primarily in CPS facing non-invasive transient/intermittent sensor anomalies where physical environment conditions change quickly. Some examples are: GPS \& radar obfuscation in tunnels, cameras \& lidars impaired by fog, cameras blinded by harsh lights, lidars affected by reflective surfaces and glass windows, infrared interfering with lidar, \textit{etc.}.
%This includes quick movement through hostile territory (\textit{e.g.} a short drive or flight through a region with a malicious attacker), confined physical attacks (\textit{e.g.} a self-driving vehicle faced with blinding lights below a bridge, reflective surfaces on the sidewalk or walls, sound or signal deflection in a tunnel, reflective glass windows of buses in traffic, \textit{etc.}) and other transient disruptive situations. 
%As described in line 17 of Algorithm \ref{fmwk_algo}, the framework will force the CPS to a safe stop once anomaly duration exceeds maximum permissible anomaly period.

%\RI{\textbf{check}. I am not sure if this needs to mentioned. It is quite a corner case. Maybe it's OK as a side here, but I think it's better if we focus on the transient anomaly class.}
%anomalies until the above condition is satisfied.
If the CPS has a reduced-functionality mode wherein it is guaranteed to not be affected by a perpetual sensor anomaly, then, rather than being brought to a safe stop after the maximum tolerable anomaly duration, the CPS can continue functioning in the reduced-functionality mode. For \textit{e.g.}, a self-driving vehicle with a primary automation layer with anomalies can use a redundant secondary layer to travel at a safe speed to the nearest repair center.

\subsection{Framework implementation}
Our framework was created with Python \& Robot Operating System (ROS) middle-ware\footnote{\url{https://github.com/kaustubhsridhar/checkpointing_and_recovery}}.
\subsection{Brief problem formulation for the case study of a differential drive ground robot}
Additional details in the problem description for the case-study are given in the Appendix (Section \ref{detailed_case}). In this Subsection, we summarize the important aspects of the problem formulation for the case study.
\begin{figure}[!t]
\centering
\includegraphics[width=\linewidth]{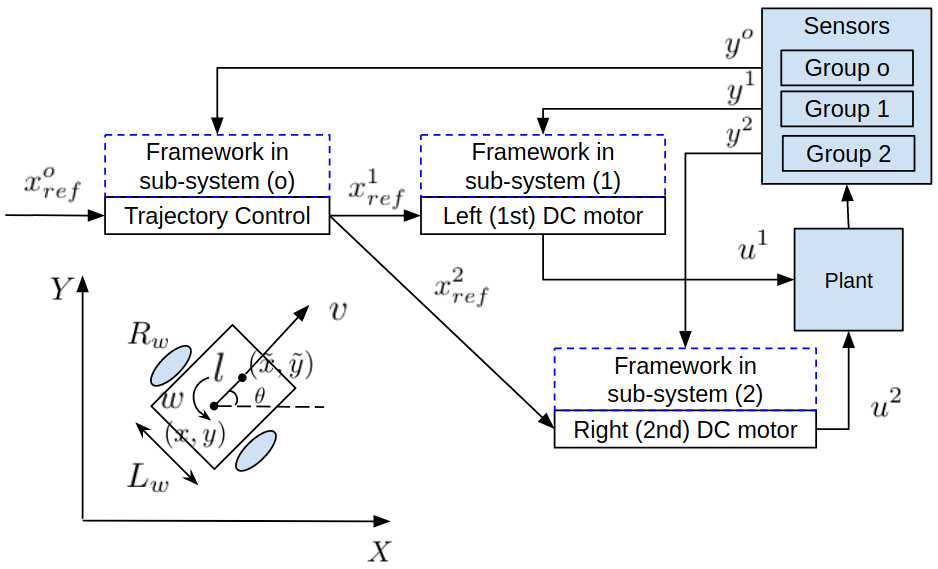}
\setlength{\abovecaptionskip}{5pt}
\setlength{\belowcaptionskip}{-5pt}
\caption{Bicycle model (bottom) \& block diagram of controllers \& framework (top-right) of a differential drive robot.}
\label{ddrive}
\end{figure}
The system and framework of the differential drive ground robot can be represented by the block diagram shown in Figure \ref{ddrive}. The ground robot has an outer trajectory control loop and two inner DC motor control loops. The outer loop provides reference values to both inner loops after transformation of outer control-inputs. The outer loop has a motion capture system. Both inner loops have encoders. The framework (and ADS) reads all sensors and provides updated state-estimates to each of the three controllers (demonstrating scalability).

The outer loop has the ground robot's trajectory controller with bicycle dynamics as follows,
\begin{align}
    x^j_{k+1} &= \left( \begin{bmatrix} v_k \cos \theta_k & v_k \sin \theta_k & w_k \end{bmatrix}^T \right) \Delta T^o + \mathcal{N}(0,Q^o)
\end{align}
with state $x^j_{k} = [x_k, y_k, \theta_k]^T$ (\textit{i.e.} $x$, $y$ position and heading) and control inputs $u^j_{k} = [v_k, w_k]^T$ (\textit{i.e.} vel. and angular vel.) as shown in Figure \ref{ddrive}. The outer loop employs sensors with the following model,
\begin{align}
    y_k^o = x_k^o + \mathcal{N}(0,R^o)
\end{align}
A dynamic inversion controller (based on first order dynamics) that considers a shifted state $l$ distance away (with $l \to 0$) is employed,
\begin{align}
        \begin{bmatrix} v_k \\ w_k \end{bmatrix} &= \begin{bmatrix} \cos \theta_k & \sin \theta_k \\ -\sin \theta_k/l & \cos \theta_k/l \end{bmatrix} \begin{bmatrix} \dot{x}_{k,ref} + k_1 (x_{k,ref} - x) \\ \dot{y}_{k,ref} + k_2 (y_{k,ref} - y_k)\end{bmatrix}
\end{align}
The inner loops have PID controllers for the left and right DC motors with dynamics and control as follows (for $j=1,2$),
\begin{align}
    x^j_{k+1} &= \left( \begin{bmatrix} -R/L & -K_{emf}/L \\ K_{tor}/J & -K_{fric}/J \end{bmatrix} x^j_k + \begin{bmatrix} 1/L \\ 0 \end{bmatrix} \mathrm{V}^j_k \right) \Delta T^j + \mathcal{N}(0,Q^j) \nonumber \\
    \mathrm{V}^j_k &= PID(w^j_{k, ref} - w^j_k)
\end{align}
with state $x^j_{k} = [i_k, w_k]^T$ (\textit{i.e.} current and angular velocity of motor) and control input $\mathrm{V}^j_{k}$ (\textit{i.e.} voltage input to motor). The inner loops have measurement model as $y_k^j = \begin{bmatrix} 0 & 1 \end{bmatrix} x_k^j + \mathcal{N}(0,R^j)$.
%\begin{align}
%    y_k^j = \begin{bmatrix} 0 & 1 \end{bmatrix} x_k^j + \mathcal{N}(0,R^j)
%\end{align}
Anomalous input of $[5,5,0]^T$ in $k=[3.25,5)$ and $-[5,5,0]^T$ in $k=[8.25,10)$ is fed to the outer loop sensors. In the same time-periods, anomalous input of $[20,000]$ and $[-20,000]$ is fed to both inner loops encoders. A specific ADS is assumed which provides list of state-estimates' elements in all three loops that depend on anomalous sensors. This ADS has detection-time of $k=[3.25,3.5) \cup [8.25,8.5)$ for all three loops. The time-periods of the outer $dt^o$ and inner loops $dt^1, dt^2$ are obtained as reciprocals of their chosen loop rates of 10 Hz and 100 Hz respectively. Checkpoints are created every second ($\mu=1$Hz).
\begin{figure}[!t]
    \centering
    \includegraphics[width=0.8\linewidth]{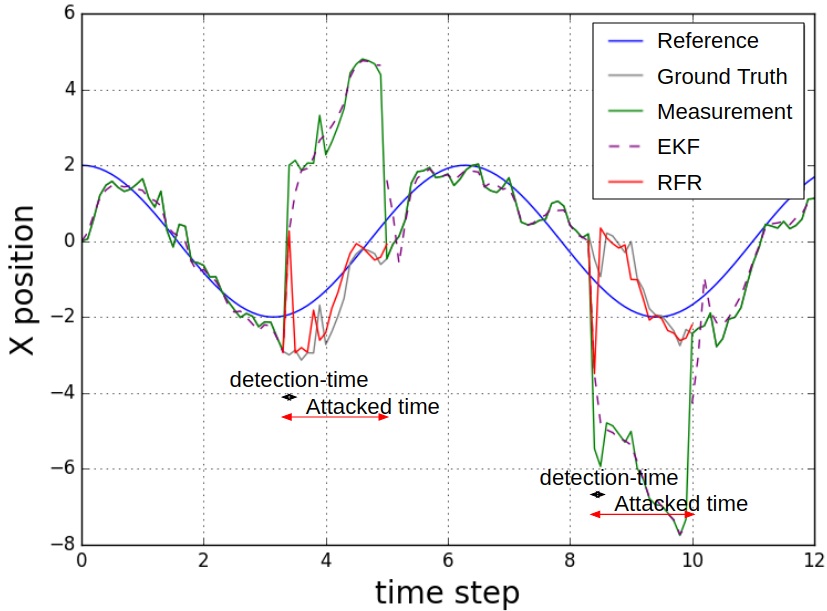}
    \setlength{\abovecaptionskip}{5pt}
    \setlength{\belowcaptionskip}{-5pt}
    \caption{Time evolution of various robot x position series'}
    \label{xw1}
\end{figure}
\begin{figure}[!t]
    \centering
    \includegraphics[width=0.8\linewidth]{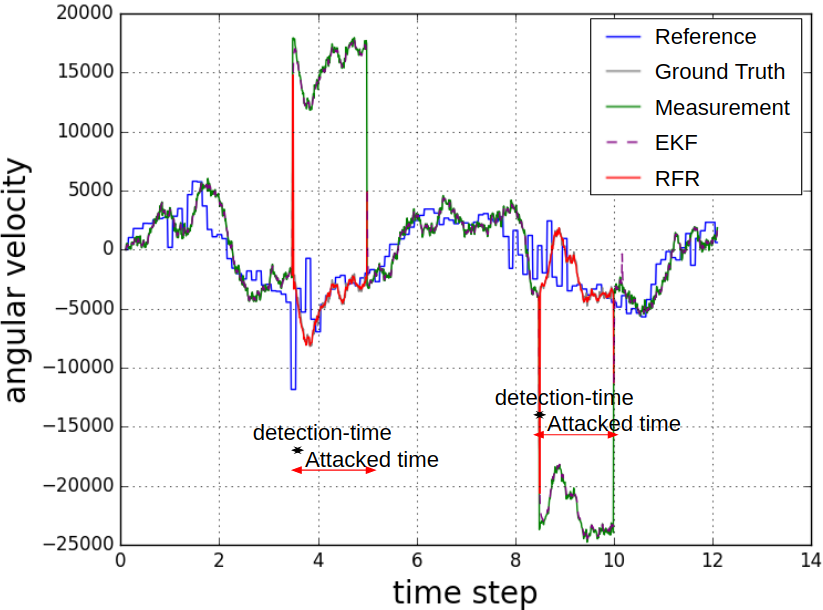}
    \setlength{\abovecaptionskip}{5pt}
    \setlength{\belowcaptionskip}{-5pt}
    \caption{Time evolution of various left DC-motor $w$ series'}
    \label{xw2}
\end{figure} 
\subsection{Simulation Results}
In this subsection, we present and explain plots generated through numerical simulation. Figure \ref{xw1} shows a plot of $x$ position of simulated ground robot vs time step and Figure \ref{xw2} shows a plot of angular velocity of left DC motor vs time step. Plots of $y$ position of robot and angular velocity of right DC motor closely resemble the previously mentioned figures and so have not been shown.

In both Figures 
%\RI{I feel like the figures could be made slightly smaller. It would look a bit more professional to not have these giant figures taking up the whole row -- but this is a personal preference, so I'd leave it up to you.}
\ref{xw1}, \ref{xw2}, we see the ground truth (GT) state tracking the reference trajectory. Controller performance is thus assured. The EKF also tracks the ground truth when there is no anomaly but swerves to the false measurements (M) in the duration of the anomaly ($k=[3.25,5) \cup [8.25,10)$). The roll-forward value (RF), on the other hand, tracks the ground truth once the anomaly is detected ($k=[3.5,5) \cup [8.5,10)$). Thus, checkpointing and roll-forward recovery is shown to produce better state-estimates than an EKF during the anomalous time period.
\begin{figure}[!t]
    \centering
    \includegraphics[width=0.75\linewidth]{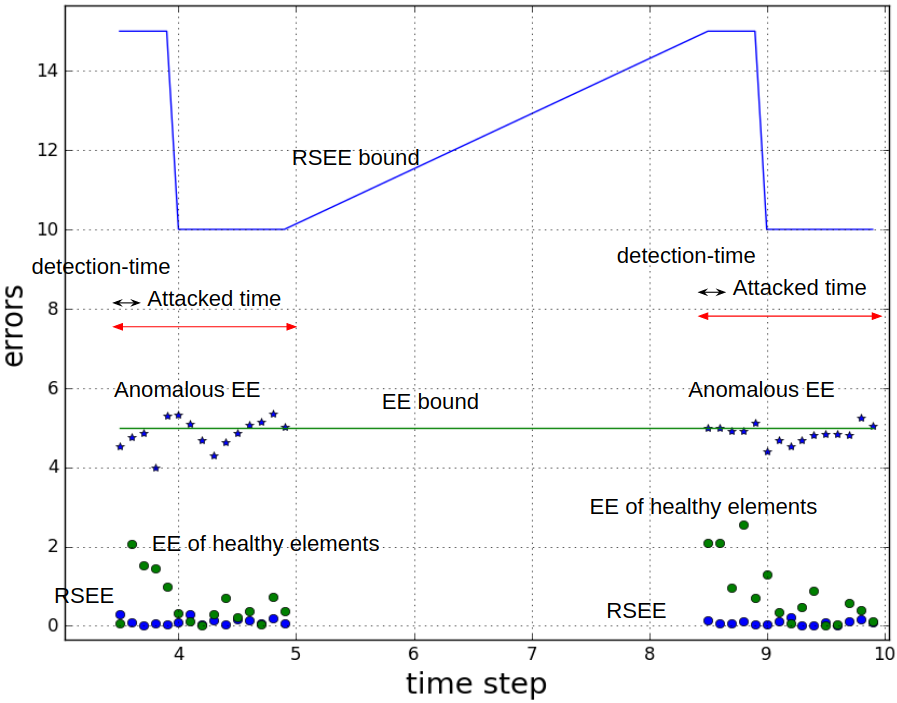}
    \setlength{\abovecaptionskip}{5pt}
    \setlength{\belowcaptionskip}{-5pt}
    \caption{Time evolution of errors \& bounds in outer loop}
    %\caption{Time evolution of recovered state estimate error (RSEE) and anomalous estimation error (anomalous EE) for anomalous x position, estimation error (EE) for healthy $\theta$ heading alongside all bounds for ground robot's outer loop}
    \label{errs1}
\end{figure} 
\begin{figure}[!t]
    \centering
    \includegraphics[width=0.775\linewidth]{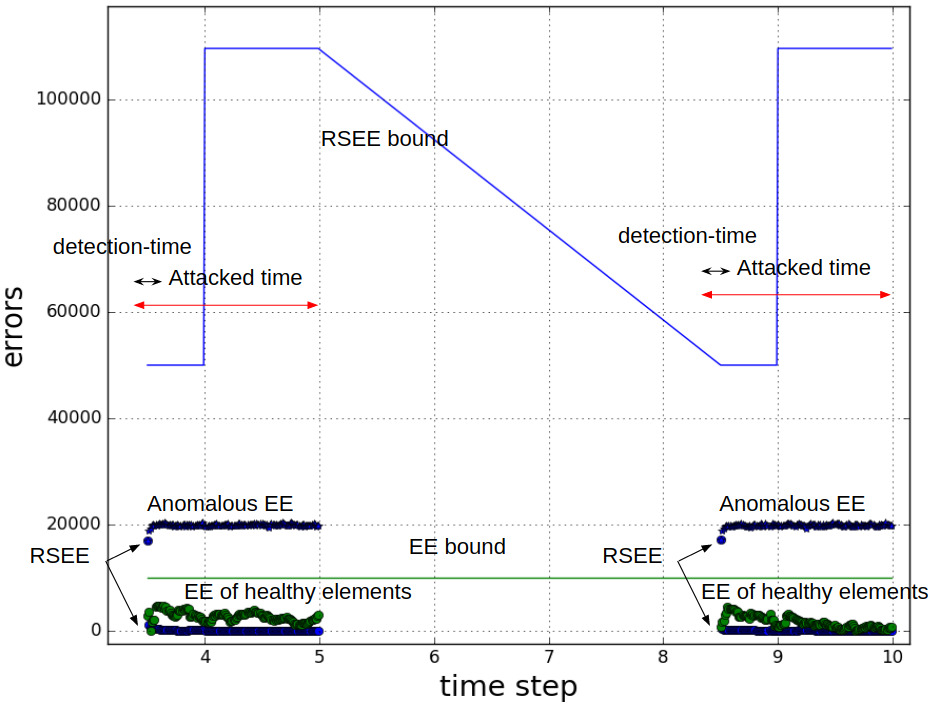}
    \setlength{\abovecaptionskip}{5pt}
    \setlength{\belowcaptionskip}{-5pt}
  \caption{Time evolution of errors, bounds in inner loop (1)}
  %\caption{Time evolution of recovered state estimate error (RSEE) and anomalous estimation error (anomalous EE) for anomalous $w$ ang. vel., estimation error (EE) for healthy $i$ current alongside all bounds for left DC motor inner loop}
  \label{errs2}
\end{figure}

In the outer loop, and in the duration of the anomaly, we have RSEE \& anomalous EKF estimate error (anomalous EE) for anomalous x position and estimation errors (EE) for healthy $\theta$ heading. This is shown in Figure \ref{errs1}. 
In the inner loops, and in the duration of the anomaly, we have RSEE \& anomalous KF estimate error (anomalous EE) for anomalous angular velocity $w$ and estimation error (EE) for healthy current $i$. This is shown for the left DC motor in Figure \ref{errs2}.
Both Figures \ref{errs1}, \ref{errs2} also show the absolute bounds of EE and RSEE from Equation \ref{ee_b} and Proposition \ref{prop_2} / Corollary \ref{lin_corollary} resp.. Both RSEE and EE errors fall comfortably within their bounds. 
%The Figure also shows the absolute bounds of EE and RSEE from Equation \ref{ee_b} and Proposition \ref{prop_2} respectively. Both RSEE and EE errors fall comfortably within their bounds. Also in the duration of the anomaly, we have estimation error (anomalous EE) between EKF estimated anomalous x position and its ground truth. This is also shown in Figure \ref{errs1}.  
%\KS{Check for perfect image scaling before submission}
%The Figure also shows the absolute bounds of RSEE and EE from Equation \ref{ee_b} and Corollary \ref{lin_bound_pred} respectively. Both RSEE and EE errors fall comfortably within their bounds. The bounds for estimation error are tighter compared to that for the recovered state-estimate error (due to the assumptions involved in bounding dynamics in Proposition \ref{prop_2}). Also in the duration of the anomaly, we have estimation error (anomalous EE) between EKF estimated anomalous angular velocity $w$ and its ground truth. This too is shown in Figure \ref{errs2}.
The anomalous EE is much larger than the RSEE for both x position and angular velocity $w$. This also shows that our framework performs better than an EKF in the presence of sensor anomalies.

%% file: 7_discussion.tex
\section{Extension to Distributed Systems} \label{discussion_sec}
In extending our work to distributed systems, we need to first ensure the reliability of communication, \textit{i.e.} no messages are lost. We have to place scrutiny on the ADS monitoring the communication mechanism which presents as an added location for anomalies. Provided we have a reliable communication protocol on which anomalies can be accurately detected, then, we will have to focus on the time-delay caused by communication.
%under communication and network attacks such as replay attacks, man-in-the-middle attacks and others \cite{network_attks}, we will have to focus on the time-delay caused by communication.
%\OS{\tetxbf{Done}. Even without attacks, we need to be sure that messages cannot be lost.} 
A simple diagrammatic argument can show that if we assume communication lag ($\nu$) is bounded by the minimum of the inverse sub-system frequencies, \textit{i.e.} $\nu \leq \min(\frac{1}{\mu^{in}}, \frac{1}{\mu^o})$, then we can directly apply our framework to distributed systems. We will leave further extension of our framework to distributed systems in non-ideal conditions to future work. 
%\OS{\textbf{Thanks}. I am not sure we need a proof in a discussion section.  Just say "a simple diagrammatic argument can show that ..."}

%% file: 8_conclusion.tex
\section{Conclusion and Future Work} \label{conclusion_sec} 
In conclusion, a framework for checkpointing and roll-forward recovery of state-estimates in nonlinear, hierarhical CPS with anomalous sensor data was proposed. The framework ensures consistent checkpointing with time-efficient roll-forward for diverse ADS in sub-systems. A tradeoff analysis between checkpointing frequency, accuracy and system resources was investigated and the related accuracy-resource gap bounded. Recovered state-estimate error bounds and maximum tolerable anomaly duration were also examined. A numerical evaluation showed scalability and better performance than a simple EKF. An extension to distributed systems under certain ideal conditions was discussed. In the future, a comprehensive extension to distributed systems can be completed.

%In the long term, we envision a future with reliable safety-critical devices that integrate this framework into the core of their software systems.

%% file: 9_appendix.tex
\section{Appendix} 
\subsection{Additional problem formulation details} \label{EKF_sec_app}
\subsubsection{\textbf{Extended Kalman Filter:}} The extended Kalman filter (EKF) is given as follows (with predict steps \eqref{3}, \eqref{4}, gain calculation \eqref{5} and update steps \eqref{6}, \eqref{7}), 
\begin{align}
    \hat{x}^j_{k+1|k} &= f^j(\hat{x}^j_{k},u^j_k), \;\;\;\;\; \hat{x}^j_{0} = x^j_{init}, \;\; P^j_{0} = \Sigma^j \label{3}\\
    P^j_{k+1|k} &= A^j_k P^j_k {A^j_k}^T + Q^j \label{4}\\
    K^j_k &= P^j_{k+1|k} {C^j_k}^T (C^j_k P^j_{k+1|k} {C^j_k}^T + R^j)^{-1} \label{5}\\
    \hat{x}^j_{k+1} &= \hat{x}^j_{k+1|k} + K^j_k(y^j_k - g^j(\hat{x}^j_{k+1|k},u^j_k)) \label{6}\\
    P^j_{k+1} &= (I - K^j_k C^j_k)P^j_{k+1|k} \;\;\;\;\; j \in \{in,o\} \label{7}
\end{align}
where $A^j_k = \left. \frac{\partial f^j}{\partial x^j} \right\vert_{\hat{x}_{k}, u_k}$ and $C^j_k = \left. \frac{\partial g^j}{\partial x^j} \right\vert_{\hat{x}_{k+1|k}, u_k}$ are Jacobians.

\subsection{Proof of Proposition \ref{prop_2}} \label{Proof_1}

For the prediction error, we have by Equation \eqref{pred_error_eq},
\begin{align*}
    e^{j}_{k+1, \ Rec} &= \left[ x^{j}_{k+1} - \hat{x}^{j,R}_{k+1} \right]_{(1,q)} \\
    &= \left[ \left( f^j(x^j_k,u^j_k) + \omega^j_{k+1} \right) -  f^j(\hat{x}^{j,R}_k,u^j_k) \right]_{(1,q)}\\
    &= \left[ \left( f^j(x^j_k,u^j_k) - f^j(\hat{x}^{j,R}_k,u^j_k) \right) + \omega^j_{k+1} \right]_{(1,q)} \\
    &= \left[ A_{k}\left(x^j_k - \hat{x}^{j,R}_k \right) + \varphi(x^j_k, \hat{x}^{j,R}_k, u^j_k) + \omega^j_{k+1} \right]_{(1,q)}\\
    &= \left[ A^j_{k} \left( f^j(x^j_{k-1},u^j_{k-1}) - f^j(\hat{x}^{j,R}_{k-1},u^j_{k-1}) \right) \right. \\
    &\left. + \omega^j_{k+1} + \varphi(x^j_k, \hat{x}^{j,R}_k, u^j_k) \right]_{(1,q)}
\end{align*}
where $\varphi(x^j_k, \hat{x}^{j,R}_k, u^j_k)$ are the nonlinear terms from Taylor series expansion of $f^j(x^j_k,u^j_k)$ about $\hat{x}^{j,R}_k$. 

By repeatedly using Taylor series expansions till checkpoint time ($k_1$), we have,
\begin{align*}
    e^{j}_{k+1, \ Rec} &= \left[ \prod_{l=k_1}^{k} A^j_{l} \delta^{j}_{k_1} +  \sum_{l=k_1}^{k} \prod_{m=l}^{k} (A^j_{m}) \omega^j_{l} \right.\\
    &\left. + \omega^j_{k+1} + \Phi(x^j_k,\hat{x}^{j,R}_k, u^j_k, ..., x^j_{k_1},\Bar{x}^{j}_{k_1}, u^j_{k_1}) \right]_{(1,q)}
\end{align*}
where $\delta^{j}_{k_1} = \left(x^j_{k_1} - \Bar{x}^{j}_{k_1} \right)$ can be obtained by a process similar to that described by Equations in \eqref{process_delta}.

$\Phi(x^j_k,..., u^j_{k_1})$ represents the accumulated nonlinear terms after repeated Taylor series expansions. Finally, with the assumptions from Equations \eqref{err_ass_0}, \eqref{err_ass_1}, \eqref{err_ass_2}, we bound the prediction error as given in Equation \eqref{pe_b}. Hence proved Proposition \ref{prop_2}.

\subsection{Proof of Proposition \ref{prop_3}} \label{Proof_3}
Since the recovered state-estimate error monotonically increases with time, it will reach the maximum permissible error $\mathbf{E}^j$ at some time $k$. We call the duration of time between the start of the attack at time $s$ and concerned time $k$, the maximum tolerable anomaly duration $T^j_{\max}$. Hence, we have $k = s + T^j_{\max}$. Then, given latest consistent checkpoint time before anomaly $k_1 = \mathcal{F}^j(s, \Delta, \mu)$, we can find $T^j_{\max}$ by solving $\mathcal{B}(k, k_1) = \mathbf{E}^j$ or equivalently the numerical minimization problem given in \eqref{T_min_prob}. This proves Proposition \ref{prop_3}.

\subsection{Proof of Proposition \ref{prop_4}} \label{Proof_4}
The accuracy-resource gap $\mathcal{P}^j$ is given as 
\begin{align}
    \mathcal{P}^j &= \hat{x}_{k,(1,q)}^{j,R} \Big\rvert_{\mu^*} - \hat{x}_{k,(1,q)}^{j,R} \Big\rvert_{\mu} \\
    &= e^j_{k, \ Rec} \Bigg\rvert_{k_1 = \ \mathcal{F}^j(s, \Delta, \mu)} - e^j_{k, \ Rec}  \Bigg\rvert_{k_1 = \ s - 1 = \ \mathcal{F}^j(s, \Delta, \mu^*)}\\
    &\leq \mathcal{B}(k, \mathcal{F}^j(s, \Delta, \mu)) - \mathcal{B}(k, s - 1) \;\;\;\;\; (\text{using } \eqref{pe_b}, \eqref{Bound})
\end{align}
which proves Proposition \ref{prop_4}.

\begin{table}[H]
    \centering
    \begin{tabular}{|p{2.2cm}|p{1.8cm}|l|l|}
        \hline 
        \row{Anomalies in sensors of subsystems $(o)$, $(in)$}{Affected anomalous values}{Solution or assumption for safe operation of CPS}\\
        \hline 
        \hline 
        \row{detected, detected}{$\hat{x}^{o}_k\to$ $u^o_k$; $\hat{x}^{in}_k\to$ $u^{in}_k$}{roll-forward recovery (RFR) in both $(o), (in)$}\\
        \hline
        \row{detected, DNE}{$\hat{x}^{o}_k\to$ $u^o_k\to$ $x^{in}_{k, ref}\to$ $u^{in}_k$}{RFR in $(o)$}\\
        \hline
        \row{DNE, detected}{$\hat{x}^{in}_k\to$ $u^{in}_k$}{RFR in $(in)$}\\
        \hline
        \row{not yet detected, DNE}{$\hat{x}^{o}_k\to$ $u^o_k\to$ $x^{in}_{k, ref}\to$ $u^{in}_k$}{assume small value of $(o)$ detection time $\mathcal{T}^{o}_k$}\\
        \hline
        \row{DNE, not yet detected}{$\hat{x}^{in}_k\to$ $u^{in}_k$}{assume small value of $(in)$ detection time $\mathcal{T}^{in}_k$}\\
        \hline
        \row{detected, not yet detected}{$\hat{x}^{o}_k\to$ $u^o_k$; $\hat{x}^{in}_k\to$ $u^{in}_k$}{RFR in $(o)$ and assume small value of $\mathcal{T}^{in}_k$}\\
        \hline
        \row{not yet detected, detected}{$\hat{x}^{o}_k\to$ $u^o_k$; $\hat{x}^{in}_k\to$ $u^{in}_k$}{RFR in (in) and assume small value of $\mathcal{T}^o_k$}\\
        \hline
        \row{not yet detected, not yet detected}{$\hat{x}^{o}_k\to$ $u^o_k$; $\hat{x}^{in}_k\to$ $u^{in}_k$}{assume small bounds on $\mathcal{T}^o_k, \mathcal{T}^{in}_k$}\\
        \hline 
        \multicolumn{3}{|p{8.2cm}|}{$^*$DNE = Do Not Exist.}\\
        \hline
    \end{tabular}
    \caption{All possible combinations of cases of sensor anomalies in the hierarchy, the values they affect and the solution or assumption that addresses them.}
    \label{table}
\end{table}

\subsection{Case-by-case analysis of anomalies in each sub-system} \label{case-by-case-analysis}
There are three cases for sensor anomalies in each sub-system: not yet detected anomalies, detected anomalies and no anomalies. The first induces detection error, the second leads to recovered state-estimate error and we have standard estimation error in all cases. We can have combinations of these three cases across the hierarchy and all such combinations are addressed by our framework. This is characterized by Table \ref{table}. It has 3 columns: the first column displays the anomaly situation in subsystems $(o), (in)$ respectively, the second column mentions which values are affected by sensor anomalies in each sub-system and the third column presents a solution or assumption for continued safe operation of the CPS. 

In the first row, both sub-systems $(o)$ and $(in)$ have detected anomalies in sensors. This affects state estimates which in turn affect control values in both sub-systems (\textit{i.e.} $\hat{x}^j_k \to u^j_k$ for $j = o, in$). For safe operation of CPS, we will have roll-forward recovery (\textit{abbreviated} RFR) in both sub-systems $(o), (in)$. In the fourth row, we have anomalies in sub-system $(o)$ that are not yet detected and anomalies in sub-system $(in)$ that do not exist (\textit{i.e.} no anomalies in $(in)$). The anomaly that hasn't been detected in $(o)$ affects its state-estimates which affects the control inputs produced by $(o)$ which in turn affects the reference values sent by $(o)$ to $(in)$ which propagates to the control value generated by $(in)$ (\textit{i.e.} $\hat{x}^{o}_k\to$ $u^o_k\to$ $x^{in}_{k, ref}\to$ $u^{in}_k$). For safe operation of CPS in this case, we assume a small detection time. The other rows in the table can be similarly interpreted.

\subsection{More details and results for case-study}
\subsubsection{\textbf{Additional details for problem formulation in the case-study of a simulated differential drive ground robot}} \label{detailed_case}

We have for the outer loop
\begin{align}
    &x^o_{ref} = \begin{bmatrix} 2 \cos(k) & 2 \sin(k) & \tan^{-1} \left( \frac{2 \sin(k) - y_k}{2 \cos(k) - x_k} \right) \end{bmatrix}^T \\
    &A^o_k = \begin{bmatrix} 0&0&-v_k \sin \theta_k \\ 0&0&v_k \cos \theta_k \\ 0&0&0 \end{bmatrix},  \Gamma^o = \begin{bmatrix} 1&0&0 \\ 0&1&0 \\ 0&0&0 \end{bmatrix}\\
    &G^{o}_k = \begin{cases} [ 1, 1, 0 ]^T, \; k \in [3.5,5) \cup [8.5, 10)\\
     [ 0, 0, 0 ]^T, \text{otherwise}
    \end{cases}
\end{align}
\begin{align}
    &x^1_{ref} = w^1_{k, ref} \in \mathbb{R}, x^2_{ref} = w^2_{k, ref} \in \mathbb{R}\\
    &T(u^o_k) = \begin{bmatrix} w^1_{k, ref} \\ w^2_{k, ref} \end{bmatrix} = \begin{bmatrix} \frac{2v - \omega L_w}{2R_w} & \frac{2v + \omega L_w}{2R_w}  \end{bmatrix}^T
\end{align}
where, at time $k$, $w^1_{k, ref}, w^2_{k, ref}$ represent reference angular vel. sent by outer loop to 1st and 2nd DC motor controllers resp., radius of wheel is $R_w = 0.05$m and length between wheels is $L_w =0.5$m. We also assume $Q^o = R^o = (0.1)^2\mathbf{I}_{3\times3}$

Next, we have for the inner loops ($j=1,2$),
\begin{align}
    G^{j}_k = \begin{cases}
      \begin{bmatrix} 0 & 1 \end{bmatrix}^T, \; k \in [3.5,5) \cup [8.5, 10)\\
      \begin{bmatrix} 0 & 0 \end{bmatrix}^T, \text{otherwise}
    \end{cases}
\end{align}
and we have resistance $R = 1 \Omega$, self-inductance $L = 0.5$H, armature, emf and viscous friction coefficients $K_{tor} = K_{emf} = K_{fric} = 0.01$ and inertial load $J = 0.01$ of the DC motor. We also have gains of the PID controller $K_p = 13.2$, $K_d = 0.275$ and $K_i = 1.525$. Lastly, we assume $Q^1 = Q^2 = (50)^2 \mathbf{I}_{2\times2}$ and $R^1 = R^2 = (50)^2$.

\begin{figure}[H]
  \centering
  \includegraphics[width=0.7\linewidth]{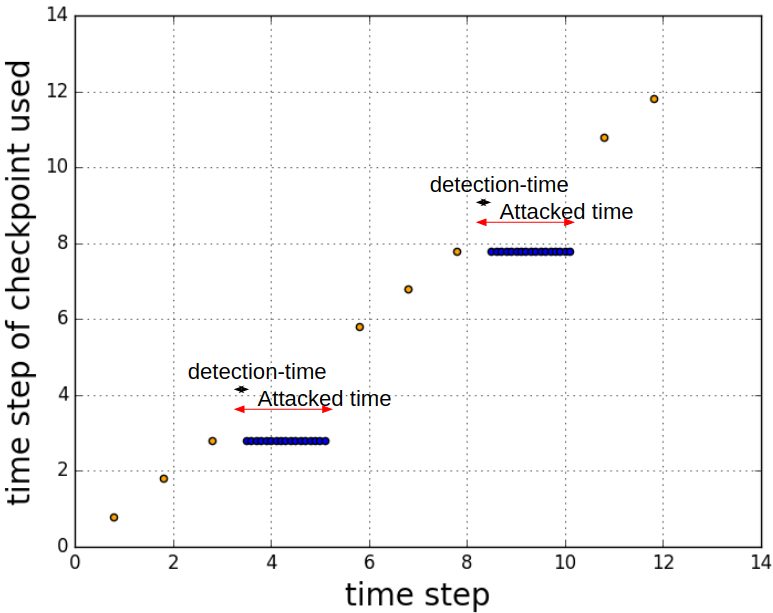}
  \setlength{\abovecaptionskip}{5pt}
  \setlength{\belowcaptionskip}{-10pt}
  \caption{Checkpoints used for roll-forward vs time step (blue) and checkpoint creation (orange) in the ground robot's trajectory control outer loop}
  \label{ckpts}
\end{figure}

\subsubsection{\textbf{Checkpointing in the case-study}}

Figure \ref{ckpts} represents the checkpointing process for the robot trajectory controller. The orange points represent checkpoints that were created at a time step given by either their x or y coordinates on the plot. The blue points represent checkpoints that were created at time step given by their y coordinate and used for roll-forward recovery at time step given by their x coordinate. Since the inner loops run faster than the outer loop, we would see more blue points in the plots for the DC motor but no other difference and so they are not shown here.